\documentclass{article}
\usepackage[final, nonatbib]{neurips_2025}

\usepackage[utf8]{inputenc} % allow utf-8 input
\usepackage[T1]{fontenc}    % use 8-bit T1 fonts
\usepackage{hyperref}       % hyperlinks
\usepackage{url}            % simple URL typesetting
\usepackage{booktabs}       % professional-quality tables
\usepackage{amsfonts}       % blackboard math symbols
\usepackage{nicefrac}       % compact symbols for 1/2, etc.
\usepackage{microtype}      % microtypography
\usepackage{xcolor}         % colors
\usepackage{wrapfig}        % wrap text around figures
\usepackage{graphicx}       % include graphics and images
\usepackage{amsmath}
\usepackage{multirow}
\usepackage[table]{xcolor}
\usepackage{stfloats}
\usepackage{enumitem}
\usepackage{subcaption}
\usepackage{bm, float}

\definecolor{best}{rgb}{0.8, 0.9, 1.0}
\definecolor{2ndbest}{rgb}{0.9, 1.0, 0.8}
\definecolor{gray_utku}{rgb}{0.4, 0.4, 0.4}
\newcommand{\best}[1]{\colorbox{best}{\textbf{#1}}}
\newcommand{\secondbest}[1]{\colorbox{2ndbest}{#1}}
\newcommand{\var}[1]{\textcolor{gray_utku}{\text{\tiny$\pm$ #1}}}
\usepackage{arydshln}

\makeatletter
\usepackage{xspace}
\def\@onedot{\ifx\@let@token.\else.\null\fi\xspace}
\DeclareRobustCommand\onedot{\futurelet\@let@token\@onedot}

\newcommand{\eqnref}[1]{Eq\onedot~\eqref{#1}}
\newcommand{\figref}[1]{Fig\onedot~\ref{#1}}

\newcommand{\secref}[1]{Section~\ref{#1}}
\newcommand{\tabref}[1]{Tab\onedot~\ref{#1}}
\newcommand{\appref}[1]{Appendix~\ref{#1}}

\def\eg{\emph{e.g}\onedot}

\title{Fast MRI for All: Bridging Access Gaps by Training without Raw Data}

\author{
  Ya\c{s}ar Utku Al\c{c}alar \\
  University of Minnesota\\
  \texttt{alcal029@umn.edu} \\
  \And
  Merve G\"{u}lle \\
  University of Minnesota\\
  \texttt{glle0001@umn.edu} \\
  \And
  Mehmet Ak\c{c}akaya\thanks{Corresponding Author} \\
  University of Minnesota\\
  \texttt{akcakaya@umn.edu}
}

\begin{document}

\maketitle

\begin{abstract}
Physics-driven deep learning (PD-DL) approaches have become popular for improved reconstruction of fast magnetic resonance imaging (MRI) scans. Though PD-DL offers higher acceleration rates than existing clinical fast MRI techniques, their use has been limited outside specialized MRI centers. A key challenge is generalization to rare pathologies or different populations, noted in multiple studies, with fine-tuning on target populations suggested for improvement. However, current approaches for PD-DL training require access to raw k-space measurements, which is typically only available at specialized MRI centers that have research agreements for such data access. This is especially an issue for rural and under-resourced areas, where commercial MRI scanners only provide access to a final reconstructed image. To tackle these challenges, we propose \textbf{C}ompressibility-inspired \textbf{U}nsupervised Learning via \textbf{P}arallel \textbf{I}maging Fi\textbf{d}elity (CUPID) for high-quality PD-DL training using only routine clinical reconstructed images exported from an MRI scanner. CUPID evaluates output quality with a compressibility-based approach while ensuring that the output stays consistent with the clinical parallel imaging reconstruction through well-designed perturbations. Our results show CUPID achieves similar quality to established PD-DL training that requires k-space data while outperforming compressed sensing (CS) and diffusion-based generative methods. We further demonstrate its effectiveness in a zero-shot training setup for retrospectively and prospectively sub-sampled acquisitions, attesting to its minimal training burden. As an approach that radically deviates from existing strategies, CUPID presents an opportunity to provide broader access to fast MRI for remote and rural populations in an attempt to reduce the obstacles associated with this expensive imaging modality. Code is available at \url{https://github.com/ualcalar17/CUPID}.
\end{abstract}

\section{Introduction} \label{sec:intro}
\begin{wrapfigure}[14]{r}{0.58\textwidth}
\begin{minipage}{0.58\textwidth}
    \centering
    \vspace{2.0em}
    \includegraphics[width=0.99\linewidth]{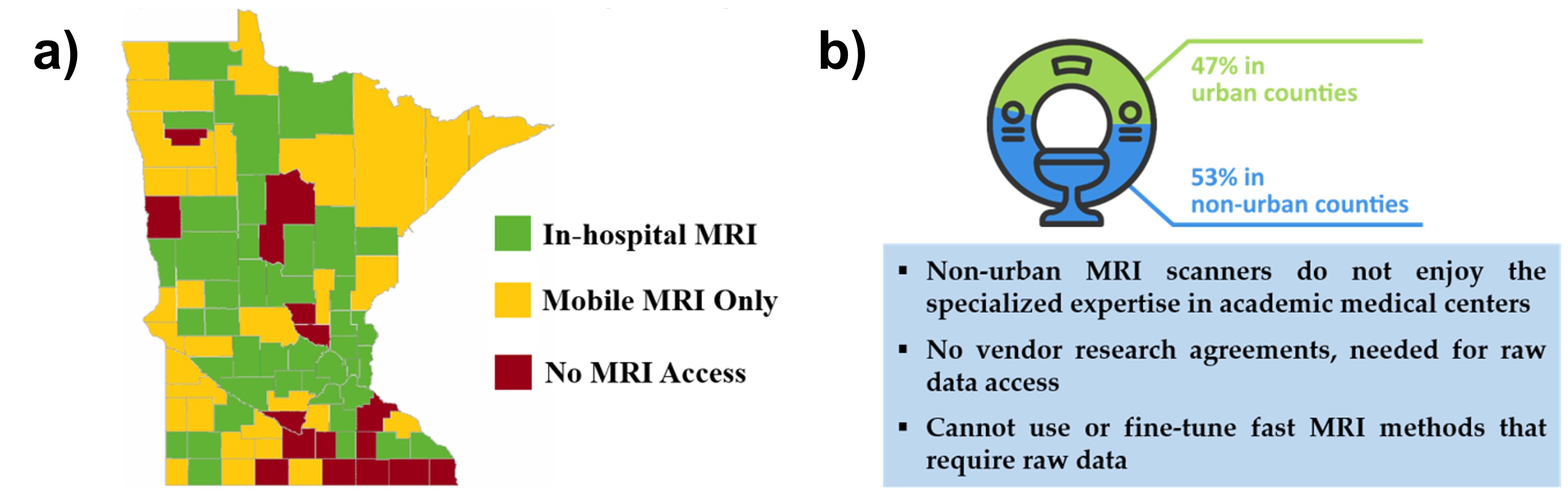}
    \caption{Many regions lack direct MRI access or rely on local/mobile units: (a) Over half of MRI services in Minnesota are in non-urban areas~\cite{burdorf2022comparing}, and (b) these scanners often lack vendor agreements for raw data access, limiting AI fine-tuning.}
  \label{fig:intro}
\end{minipage}
\end{wrapfigure}

Magnetic resonance imaging (MRI) is a central tool in modern medicine, offering multiple soft tissue contrasts and high diagnostic sensitivity for numerous diseases. However, MRI is among the most expensive medical imaging modalities, in part due to its long scan times. Demand for MRI scans has shown an annual growth rate of 2.5\%, while the number of MRI units per capita has increased by 1.8\% in a similar time frame~\cite{martella2023diagnostic-MRIdemand}. This mismatch has further increased the wait times for MRI exams~\cite{bartsch2024advanced-MRIdelays,hofmann2023-MRIdelays}, particularly in rural areas and under-resourced/remote communities~\cite{burdorf2022comparing}, shown in \figref{fig:intro}. Thus, techniques for fast MRI scans that can reduce overall scan times without compromising diagnostic quality~\cite{akcakaya2022reconbook} are critical for improving the throughput of MRI. Computational MRI approaches, including partial Fourier imaging ~\cite{mcgibney1993partialFourier}, parallel imaging~\cite{pruessmann1999sense,griswold2002grappa}, compressed sensing (CS)~\cite{lustig2007sparse}, and more recently deep learning (DL)~\cite{hammernik2018VarNet,schlemper2018deep} have been developed for accelerating MRI.

In most MRI centers, parallel imaging remains the most widely used approach for the reconstruction of routine clinical images. The acceleration rates afforded by these methods, however, are limited due to noise amplification and aliasing artifacts. DL-based methods, especially physics-driven DL (PD-DL) approaches, offer state-of-the-art improvements over parallel imaging~\cite{knoll2020deep-survey}. However, the translation of PD-DL to clinic has been hindered by generalizability and artifact issues related to details not well represented in training databases, in other words when faced with out-of-distribution samples at test time~\cite{eldar2017challenges-SPM,knoll2020fastmri_dataset-journal,muckley2021fastMRIchallenge-2,antun2020instabilities}. This is a problem for many typical MRI centers, whose population characteristics do not necessarily align with specialized MRI centers in urban settings, where training databases for PD-DL are currently curated.

In such cases, fine-tuning of the PD-DL model on the target population may be beneficial~\cite{knoll2019assessment,dar2020transfer,yaman2022zeroshot,chandler2023overcoming}. However, a major roadblock for this strategy is that all current training methods for PD-DL require access to raw MRI data. Such access requires research agreements with MR vendors, and is typically not available outside specialized/academic MRI centers. This is especially an issue for rural and under-resourced areas, where commercial MRI scanners only provide access to a final image, reconstructed via parallel imaging.

In this work, we tackle these challenges associated with typical PD-DL training, and propose \textbf{C}ompressibility-inspired \textbf{U}nsupervised Learning via \textbf{P}arallel \textbf{I}maging Fi\textbf{d}elity (CUPID), which trains PD-DL reconstruction from routine clinical images, \eg, in Digital Imaging and Communications in Medicine (DICOM) format. Succinctly, CUPID uses a compressibility-inspired term to evaluate the goodness of the output, while ensuring the output is consistent with parallel imaging via well-designed input perturbations. CUPID can be used both with database-training and in a subject-specific/zero-shot manner, attesting to its minimal fine-tuning burden. Our key contributions include:
\begin{itemize}[leftmargin=*, itemsep=0em, topsep=0pt]
\item We introduce CUPID, a novel method that enables high-quality training of PD-DL reconstruction in unsupervised and zero-shot/subject-specific settings. By using only routine clinical reconstructed MR images, CUPID eliminates the need for access to raw k-space measurements. To the best of our knowledge, our method is the first attempt to train PD-DL networks using only these images that are exported from the scanner.
\item CUPID is trained on DICOM images acquired at the \emph{target acceleration rate}, which often have noise and aliasing artifacts due to high sub-sampling, in an unsupervised manner. Note this is a deviation from other methods that use reference fully-sampled DICOM images to train a likelihood or score function, such as generative models.
\item CUPID uses a novel unsupervised loss formulation that enforces fidelity with using parallel imaging algorithms via carefully designed perturbations, in addition to evaluating the compressibility of the output image. This parallel imaging fidelity ensures the network does not converge to overly sparse solutions.
\item We provide a comprehensive evaluation encompassing both retrospective and prospective acquisitions at target acceleration rates, along with expert radiologist assessments and pathology cases from FastMRI+~\cite{zhao2022fastmri_plus}, demonstrating that CUPID achieves reconstruction quality comparable to leading supervised and self-supervised methods that require raw k-space data, while outperforming conventional CS techniques and diffusion-based generative methods.

\end{itemize}

\section{Background and Related Work} \label{sec:background}

%-------------------------------------------------------------------------
\subsection{MRI forward model and conventional methods for MRI reconstruction}
MRI raw data is acquired in the frequency domain of the image, referred to as k-space. For fast MRI, data is acquired in the sub-Nyquist regime by undersampling the acquisition in k-space. In this case, the forward acquisition model relating the image ${\bf x} \in {\mathbb C}^n$ to these raw MRI data (or k-space) measurements is given as:
\begin{equation}
    \mathbf{y}_\Omega = \mathbf{E}_\Omega\bf{x} + \bf{n},
\end{equation}
where $\mathbf{y}_{\Omega}$ denotes the acquired k-space data corresponding to the undersampling pattern $\Omega$ with $|\Omega| = m < n$. $\mathbf{E}_\Omega$ denotes the multi-coil encoding operator that includes information from $n_c$ receiver coils, each of which are sensitive to a different part of the image~\cite{hamilton2017recent}. When the acceleration rate $R = n/m$ is less than $n_c$, this system of equations is over-determined due to the redundancies among the receiver coils. Parallel imaging uses these redundancies to solve the maximum likelihood estimation problem under i.i.d. Gaussian noise~\cite{pruessmann1999sense}:
\begin{equation} 
{\bf x}_\textrm{PI} = \arg\min_{\bf x} \|\mathbf{y}_\Omega - \mathbf{E}_\Omega \mathbf{x}\|_2^2  = ({\bf E}_\Omega^H{\bf E}_\Omega)^{-1} {\bf E}_\Omega^H {\bf y}_\Omega. \label{eq:leastsq_PI} \end{equation}
Numerically, this can be solved directly for certain undersampling patterns~\cite{pruessmann1999sense} or more broadly iteratively using conjugate gradient (CG)~\cite{pruessmann2001cgsense}. Using the equivalence of multiplication in image domain and convolutions in k-space~\cite{uecker2014espirit}, it can also be solved as an interpolation problem in k-space~\cite{griswold2002grappa}. 
Parallel imaging remains the most clinically used acceleration method for MRI, with some MR systems using the image-based reconstruction, while others utilizing the equivalent k-space interpolation.

In modern computational MRI, additional regularization is often incorporated into the objective function~\cite{hammernik2023SPM}:
\begin{equation} \arg\min_{\bf x} \|\mathbf{y}_\Omega - \mathbf{E}_\Omega \mathbf{x}\|_2^2 + \mathcal{R}(\mathbf{x}), \label{eq:least_sq_mri} \end{equation}
where $\mathcal{R}(\cdot)$ denotes a regularizer. For instance, CS uses the idea that images should be compressible in an appropriate transform domain~\cite{lustig2007sparse}, and uses $\mathcal{R}(\mathbf{x}) = \tau||{\bf Wx}||_1$, where $\tau$ is the regularization weight, ${\bf W}$ is a linear sparsifying transform such as a discrete wavelet transform (DWT) and $||\cdot||_1$ is the $\ell_1$ norm.

%-------------------------------------------------------------------------
\subsection{PD-DL reconstruction via algorithm unrolling}
Among different PD-DL methods~\cite{ahmad2020plug, gilton2021deep, knoll2020deep-survey}, unrolled networks~\cite{monga2021algorithm} remain the highest performer in reconstruction challenges~\cite{hammernik2023SPM, muckley2021fastMRIchallenge-2}. These methods unroll iterative algorithms for solving the regularized least squares objective in \eqnref{eq:least_sq_mri}~\cite{fessler2020SPM}, such as proximal gradient descent~\cite{schlemper2018deep, hosseini2020dense} or variable splitting with quadratic penalty (VS-QP)~\cite{aggarwal2019MoDL}, over a fixed number of steps. Unrolled networks are conventionally trained using supervised learning over a database, where the reference raw k-space measurements are first retrospectively undersampled to form ${\bf y}_{\Omega}$. Subsequently, the network is trained to map to the original full reference k-space or the corresponding reference image~\cite{hammernik2018VarNet, aggarwal2019MoDL,junno2025_TE-MRI_NIPS} by minimizing:
\begin{equation}
    \min_{\boldsymbol{\theta}} \mathbb{E} \left[ \mathcal{L} \left( \mathbf{y}_{\text{ref}}, \mathbf{E}_{\text{full}}(f(\mathbf{y}_{\Omega}, \mathbf{E}_{\Omega}; \boldsymbol{\theta})) \right) \right]
\end{equation}
where $\boldsymbol{\theta}$ are the network parameters, $f(\mathbf{y}_{\Omega}, \mathbf{E}_{\Omega}; \boldsymbol{\theta})$ denotes the network output for inputs ${\bf y}_{\Omega}$ and ${\bf E}_{\Omega}$, $\mathbf{E}_{\text{full}}$ is the fully-sampled encoding operator, $\mathbf{y}_{\text{ref}}$ is the fully-sampled reference k-space data, and ${\cal L}(\cdot,\cdot)$ is a loss function.

%-------------------------------------------------------------------------
\subsection{Self-supervised and unsupervised methods}
Obtaining fully-sampled reference data in MRI can be infeasible due to prolonged scan durations, organ movement in acquisitions such as real-time cardiac imaging or myocardial perfusion~\cite{leiner2023cardiacMRI}, or signal decay in acquisitions like diffusion MRI with EPI~\cite{ugurbil2013pushing}. To enable training of PD-DL networks without fully sampled raw MRI data, a variety of unsupervised learning methodologies have emerged~\cite{akcakaya2022_SPMsurvey}, including self-supervised learning techniques~\cite{yaman2020SSDU,chen2021equivariant} and generative modeling approaches~\cite{jalal2021robust,chung2022scoreMRI,chung2024decomposed}.

Self-supervised methods generate supervisory labels from the undersampled data itself~\cite{yaman2020SSDU,chen2021equivariant,millard2023theoretical,hu2024spicer}. A pioneering method in this field, self-supervision via data undersampling (SSDU)~\cite{yaman2020SSDU,yaman2022mmssdu}, involves partitioning the acquired measurement indices $\Omega$ into two disjoint subsets ($\Omega = \Lambda \cup \Theta$) to train the network in a self-supervised manner:
\begin{equation}
    \min_{\boldsymbol{\theta}} \mathbb{E} \left[ \mathcal{L} \left( \mathbf{y}_{\Lambda}, \mathbf{E}_{\Lambda}(f(\mathbf{y}_{\Theta}, \mathbf{E}_{\Theta}; \boldsymbol{\theta})) \right) \right]
\end{equation}
Even though these self-supervision based approaches demonstrate exceptional performance across various tasks, they lack the ability to train the model without access to undersampled raw data, as they cannot operate solely using images that are exported from the scanner.

Conversely, generative methods learn the prior distribution of the given dataset, which is then leveraged in conjunction with a log-likelihood data term during the testing phase. Although recent methods based on diffusion/score-based models have shown substantial promise, these methods require large amounts of high-quality images either reconstructed from raw data~\cite{jalal2021robust,luo2023bayesian} or as DICOMs~\cite{chung2022scoreMRI}, as well as  computational resources to perform the training, both of which  may not be feasible in the setups we are focused on.

\section{Methodology} \label{sec:methods}
%-------------------------------------------------------------------------
\subsection{Why is it important to train PD-DL networks without raw k-space data?} \label{sec:why_CUPID}

A major limitation in the clinical adoption of deep learning (DL)-based MRI reconstruction is the issue of generalizability~\cite{hammernik2018VarNet,johnson2021evaluation,knoll2019assessment,muckley2021fastMRIchallenge-2,darestani2021measuring,heckel2024deep,zhao2024whole}. Models trained on data from a specific scanner, patient population, or acquisition protocol often fail when applied to different settings due to distribution shifts~\cite{muckley2021fastMRIchallenge-2,kustner2024predictive}. This lack of robustness can lead to artifacts or hallucinations which are false structures that resemble real anatomy, but are not present in the underlying data~\cite{knoll2019assessment,muckley2021fastMRIchallenge-2,heckel2024deep,kustner2024predictive}.

Studies have shown that pre-trained models struggle to adapt to real-world variations, particularly in centers where patient demographics, scanner configurations, or imaging protocols differ from the original training data~\cite{heckel2024deep}. This limitation was highlighted in the FastMRI challenges in the transfer track~\cite{muckley2021fastMRIchallenge-2}, where networks trained on images from one vendor failed to generalize to data from another. Traditional solutions involve fine-tuning models with additional raw k-space data from the target domain~\cite{darestani2021accelerated,yaman2022zeroshot}, but this is infeasible in many clinical environments. While all MRI scanners inherently sample k-space data, the majority of scanners outside specialized academic or research institutions (\eg, as those in local hospitals and mobile MRI units) lack the ability to export this data due to vendor restrictions~\cite{winter2024open}. As a result, most DL-based reconstruction methods, which typically rely on raw k-space during training or fine-tuning, cannot be applied in these settings.

Outside the context of PD-DL methods, diffusion models as generative priors offer an alternative in this setting. Initial promising results~\cite{jalal2021robust,chung2022scoreMRI,song2022solving-medical} have suggested more robustness to sampling pattern shifts, and some robustness to anatomy changes, though more recent works have argued that diffusion priors also face generalizability issues for out-of-distribution samples~\cite{barbano2025steerable}. Thus, while promising, the robustness of the learned prior to distribution shifts, as well as the heuristic tuning of data fidelity parameters during posterior sampling, require further investigation. Furthermore, diffusion models are trained on either raw data \cite{jalal2021robust,aali2025ambient-diff}, fully-sampled magnitude~\cite{chung2022scoreMRI} or complex-valued~\cite{chung2024decomposed,alcalar2025_CAMSAP_ZADS,gulle2025consistency} DICOMs. However, they have not been trained on sub-sampled DICOMs that still contain artifacts, as explored in this work. Finally, there are other challenges for the clinical applicability and acceptance of diffusion models in MRI, including the longer inference times compared to PD-DL, and more importantly the stochasticity of the output.

In light of these, training/fine-tuning PD-DL networks without raw k-space data is not just a matter of convenience, but a necessary step toward ensuring that DL-based MRI reconstruction can generalize across diverse clinical settings.

\begin{figure}[t]
    \centering
    \includegraphics[width=\textwidth]{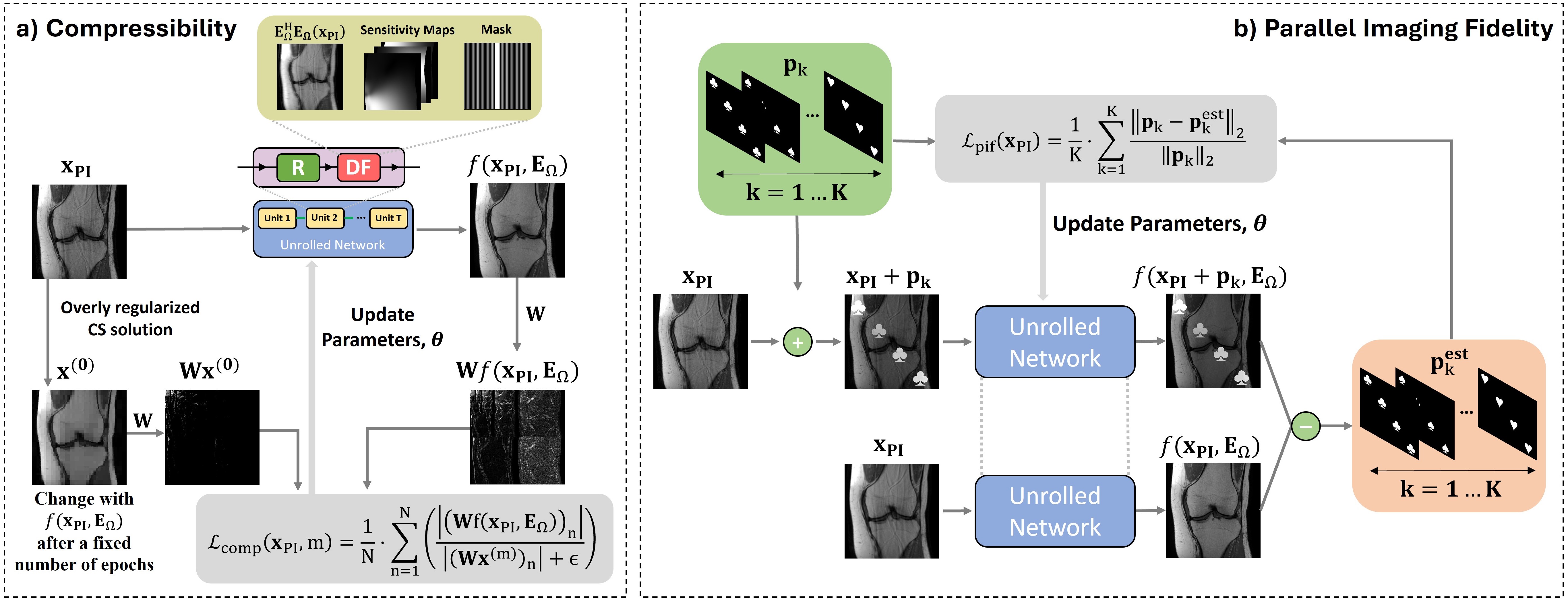}
    \caption{Our Compressibility-inspired Unsupervised Learning via Parallel Imaging Fidelity (CUPID) method trains PD-DL models in an unsupervised and/or zero-shot manner without requiring any raw k-space data. The network is unrolled for $T$ units, with each unit consisting of regularizer (R) and data fidelity (DF). The proposed loss function comprises two terms: (a) a reweighted $\ell_1$ component that assesses the compressibility of the network's output; (b) a fidelity term that ensures the output stays consistent with parallel imaging reconstructions via carefully designed perturbations, thereby preventing the network from producing a sparse all-zeros output.}
    \label{Fig:method_fig}
\end{figure}

%-------------------------------------------------------------------------
\subsection{Training without raw k-space data}
In this study, we introduce a novel framework to train PD-DL models, utilizing only  routinely available clinical images exported directly from MRI scanners. 
Recently, inspired by the connections between PD-DL and compressibility-based processing~\cite{gu2022revisiting}, a compressibility-inspired loss was proposed to evaluate the goodness of unsupervised PD-DL training~\cite{alcalar2024_ISBI}. However, this approach still requires access to raw k-space data to stabilize training, making it unsuitable for our goals. Here, we adapt the compressibility idea and augment it with a parallel imaging fidelity to successfully reconstruct clinical images in DICOM format without needing any raw k-space data.

\paragraph{Compressibility aspect of the loss formulation} Compressibility/sparsity in the output of the PD-DL network can be enforced by utilizing a weighted $\ell_1$ norm \cite{alcalar2024_ISBI}, which has been demonstrated to provide a closer approximation to the $\ell_0$ norm compared to the standard $\ell_1$ norm~\cite{candes2008enhancing}. Thus, this compressibility of the output image in CUPID is achieved by the loss term
\begin{equation}
   \mathcal{L}_{\text{comp}}( \mathbf{x}_\text{PI}, \text{m}) = \frac{1}{N} \cdot \sum_{n=1}^N \left( \frac{|(\mathbf{W}f(\mathbf{x}_\text{PI},\mathbf{E}_\Omega))_n|}{|(\mathbf{W}\mathbf{x}^{(\text{m})})_n|+\epsilon} \right), \label{eq:loss1}
\end{equation}
where $\mathbf{x}_\text{PI}$ denotes the DICOM input acquired using parallel imaging, $\bf{W}$ represents the wavelet transform, $N$ is the total number of wavelet coefficients and $\bf{x}^{(\text{m})}$ signifies the signal estimate following the training during the $m^{\text{th}}$ reweighting step. Similar to~\cite{alcalar2024_ISBI, candes2008enhancing}, we chose the initial weights from a CS reconstruction that has a large regularization and $\epsilon$ is added for numerical stability. Note, here we redefined $f(\cdot,\cdot)$ without the network parameters, $\boldsymbol{\theta}$, and used $\mathbf{x}_\text{PI}$ as the network input instead of $\bf{y}_\Omega$, to simplify notation.

\paragraph{Parallel imaging fidelity via perturbation equivariance} Relying solely on \eqnref{eq:loss1} will result in inaccurate training as the neural network learns to produce an all-zeros image in an effort to drive the wavelet coefficients in the numerator to zero, which minimizes the loss function in \eqnref{eq:loss1}. In~\cite{alcalar2024_ISBI}, fidelity with raw k-space data was used to avoid this training issue. In our setting without raw k-space access, we introduce a novel fidelity operator that stabilizes the training of the reconstruction algorithm, building on ideas from parallel imaging.
Specifically, we define a secondary loss term that enforces parallel imaging fidelity formulated as:
\begin{equation}
   \mathcal{L}_{\text{pif}}( \mathbf{x}_\text{PI}) =  %\min_{\boldsymbol{\theta}} 
   \mathbb{E}_{\bf p} \big[{||\mathbf{p} - \mathbf{p}^\text{est}||_2}\big/{||\mathbf{p}||_2}\big], \label{eq:loss2}
\end{equation}
where $\mathbf{p}^\text{est} = f(\mathbf{x}_\text{PI}+\mathbf{p},\mathbf{E}_\Omega) - f(\mathbf{x}_\text{PI},\mathbf{E}_\Omega)$. To motivate this loss term, we build upon the theoretical foundation of equivariant imaging (EI)~\cite{chen2021equivariant}. In EI, an equivariance property for the composition of the forward operator and the reconstruction network is assumed with respect to a transformation group, and necessary conditions are derived for signal recovery \cite{chen2021equivariant}.

Our work adapts this framework by introducing a novel transformation group 
tailored to parallel imaging (PI) in MRI. We note that parallel imaging reconstructions exhibit an equivariance property with respect to a specific group of perturbations ${\cal P} = \{\mathbf{p}_k\}_{k=1}^K$, which are designed to ensure that their $R$-fold aliasing do not create overlaps in the field-of-view. For ${\bf p} \in {\cal P}$, we let $T_{\bf p}(\mathbf{x}) = \mathbf{x} + \mathbf{p}$ be the affine perturbation from this group. Our parallel imaging fidelity then assumes equivariance of the reconstruction network to perturbations from this group as follows:
\begin{equation}
    f(T_{\bf p}\big(\mathbf{x}_{\text{PI}}), {\bf E}_\Omega\big) = f(\mathbf{x}_{\text{PI}} + \mathbf{p}, {\bf E}_\Omega) = f(\mathbf{x}_{\text{PI}},{\bf E}_\Omega) + {\bf p}= T_{\bf p}\big(f(\mathbf{x}_{\text{PI}},{\bf E}_\Omega) \big),
\end{equation}
Our second loss term, defined in \eqnref{eq:loss2}, promotes this equivariant behavior by penalizing deviations from this property. We also note this set of perturbations satisfy the necessary conditions in \cite{chen2021equivariant}. In doing so, it implicitly encodes parallel imaging consistency without relying on access to raw k-space data. Our final loss function for CUPID is:
\begin{equation}
   \mathcal{L}_{\text{CUPID}} = \mathcal{L}_{\text{comp}} + \lambda \cdot \mathcal{L}_{\text{pif}}, \label{eq:loss_full}
\end{equation}
where $\lambda$ is a trade-off parameter between two terms.

\paragraph{Perturbation design specifics} The perturbations used for $R$-fold acceleration are constructed to avoid aliasing overlaps in the field-of-view, ensuring they are resolvable via parallel imaging. This design aligns with the equivariance assumption introduced above: the network should recover an accurate estimate of $\mathbf{x}$ from $\mathbf{x}_{\text{PI}}$, and $\mathbf{x} + {\bf p}$ from $\mathbf{x}_{\text{PI}} + {\bf p}$. Both cases are illustrated in \figref{Fig:method_fig}.

From an implementation perspective, the expectation over ${\bf p}$ is calculated over $K$ such perturbations $\{{\bf p}_k\}$. The fold-over constraint for each $\{{\bf p}_k\}$ is achieved by picking the perturbations as randomly rotated and positioned letters, numbers, card suits or other shapes that have different intensity values. These choices also ensure that high-frequency information, such as edges, are accurately reconstructed by the regularization process. Further information on this is given in \appref{appx:pert_design}.

\paragraph{Subject-specific/zero-shot application} In resource-limited settings, it may be more practical to fine-tune the method using only a few subjects, or even a single subject, to significantly reduce computational costs. As \eqnref{eq:loss1} does not solely focus on the subtraction between two entities, it lacks an inherent mechanism to drive the loss to zero through overfitting. Therefore, CUPID can be tailored to suit a scan-specific context~\cite{akcakaya2019RAKI} without any modification to the loss given in \eqnref{eq:loss_full}.

\section{Evaluation}
%-------------------------------------------------------------------------
\subsection{Experimental setup and implementation details} \label{sec:exp_setup}
We conducted a thorough evaluation of our method, assessing its performance through both qualitative and quantitative analyses, and focused on uniform/equidistant patterns which produces coherent artifacts that are more difficult to remove compared to the incoherent artifacts from random undersampling~\cite{knoll2019assessment}.

\paragraph{Retrospective undersampling setup} In our retrospective studies, we used fully-sampled multi-coil knee and brain MRI data from the fastMRI database, acquired with relevant institutional review board approvals~\cite{knoll2020fastmri_dataset-journal,zbontar2019fastmri_dataset-arXiv}. Knee dataset included fully-sampled coronal proton density-weighted (coronal PD) and PD with fat suppression (coronal PD-FS) data, both with matrix sizes of $320\times320$. For brain MRI, axial T2-weighted (ax T2) and axial FLAIR (ax FLAIR) datasets with matrix size of $320\times320$ are used. Both knee datasets comprise data collected from 15 receiver coils, while the brain datasets include data collected from 16 and 20 receiver coils for the ax T2 and FLAIR, respectively.

Every dataset was retrospectively undersampled using a uniform/equidistant pattern at $R=4$. 24 lines of auto-calibration signal (ACS) from center of the raw k-space data were kept. DICOM images to train our proposed model were reconstructed using parallel imaging (CG-SENSE), solving $\mathbf{x}_\text{PI} = (\mathbf{E}_\Omega^H\mathbf{E}_\Omega)^{-1}\mathbf{E}_\Omega^H\mathbf{y}_\Omega$. For each dataset, models were trained using $300$ slices, and testing was performed using $380$ slices for knee MRI and $300$ slices for brain MRI, from distinct subjects. More details about the implementation of the PD-DL models are provided in \appref{appx:imp_details}.

\paragraph{Prospective undersampling setup} In this experiment, we replicate the practical pipeline for CUPID, where data is acquired at the desired high acceleration rate, and reconstructed to ${\bf x}_\textrm{PI}$ with noise and aliasing artifacts, using parallel imaging. To this end, a 3D MP2RAGE sequence~\cite{marques2010mp2rage} was acquired on a 7T Siemens Magnetom MRI scanner, with institutional review board approval, and matrix size$= 320 \times 300 \times 224$, isotropic resolution $=0.75\textrm{mm}$, prospective undersampling $R=4$ (in $k_y$ only). This is the desired target acceleration, which is higher than the standard clinical acquisition. Low-resolution images were acquired in the same orientation for sensitivity estimation~\cite{krueger2023rapid}. Training and reconstruction with CUPID was done in a zero-shot subject-specific manner.

\subsection{Comparison methods} \label{sec:comparison_methods}
We compared our method with several database training methods that have access to raw k-space data, including supervised PD-DL~\cite{hammernik2018VarNet, aggarwal2019MoDL, knoll2020deep-survey}, SSDU~\cite{yaman2020SSDU}, and equivariant imaging (EI)~\cite{chen2021equivariant}. All PD-DL methods used the same unrolled network and components (\appref{appx:imp_details}) to ensure that only the training process differed for fair comparisons. We also compared CUPID with recent diffusion model based techniques, ScoreMRI~\cite{chung2022scoreMRI} and DDS~\cite{chung2024decomposed}. These train a time-dependent score function using denoising score matching on a large dataset of reference fully-sampled DICOM images or raw data, and uses this score function during inference to sample from the conditional distribution given the measurements. Finally, we also included conventional reconstruction methods, CG-SENSE, which was used to generate the original ${\bf x}_\textrm{PI}$ as the clinical baseline comparison that is typically not clinically usable at high acceleration rates~\cite{hammernik2018VarNet}, as well as CS~\cite{lustig2007sparse}.

We note that all aforementioned methods use $\mathbf{E}_\Omega^H\mathbf{y}_\Omega$ for data fidelity during inference, since this is what is needed to run gradient descent or conjugate gradient on $\ell_2$-based fidelity terms including ${\bf y}_{\Omega}$ and ${\bf E}_{\Omega}{\bf x}$, as seen in \eqnref{eq:leastsq_PI} or SuppMat Sec. 6.
$\mathbf{E}_\Omega^H\mathbf{y}_\Omega$ can be formed without accessing ${\bf y}_{\Omega}$ by simply multiplying $\mathbf{x}_\text{PI}$ with $\mathbf{E}_\Omega^H\mathbf{E}_\Omega$. Note that ${\bf E}_{\Omega}$ includes information about the undersampling pattern $\Omega$, which is completely known from the acquisition parameters, and coil sensitivities, which can be estimated from separate calibration scans in DICOM format~\cite{krueger2023rapid}. We emphasize that what sets CUPID apart from other PD-DL strategies is that it is the only one that can train the unrolled network without using ${\bf y}_\Omega$. Thus, without loss of generality, ${\bf E}_{\Omega}$ is known both in training and in testing for all methods.

\begin{figure}
    \centering
    \includegraphics[width=1.0\textwidth]{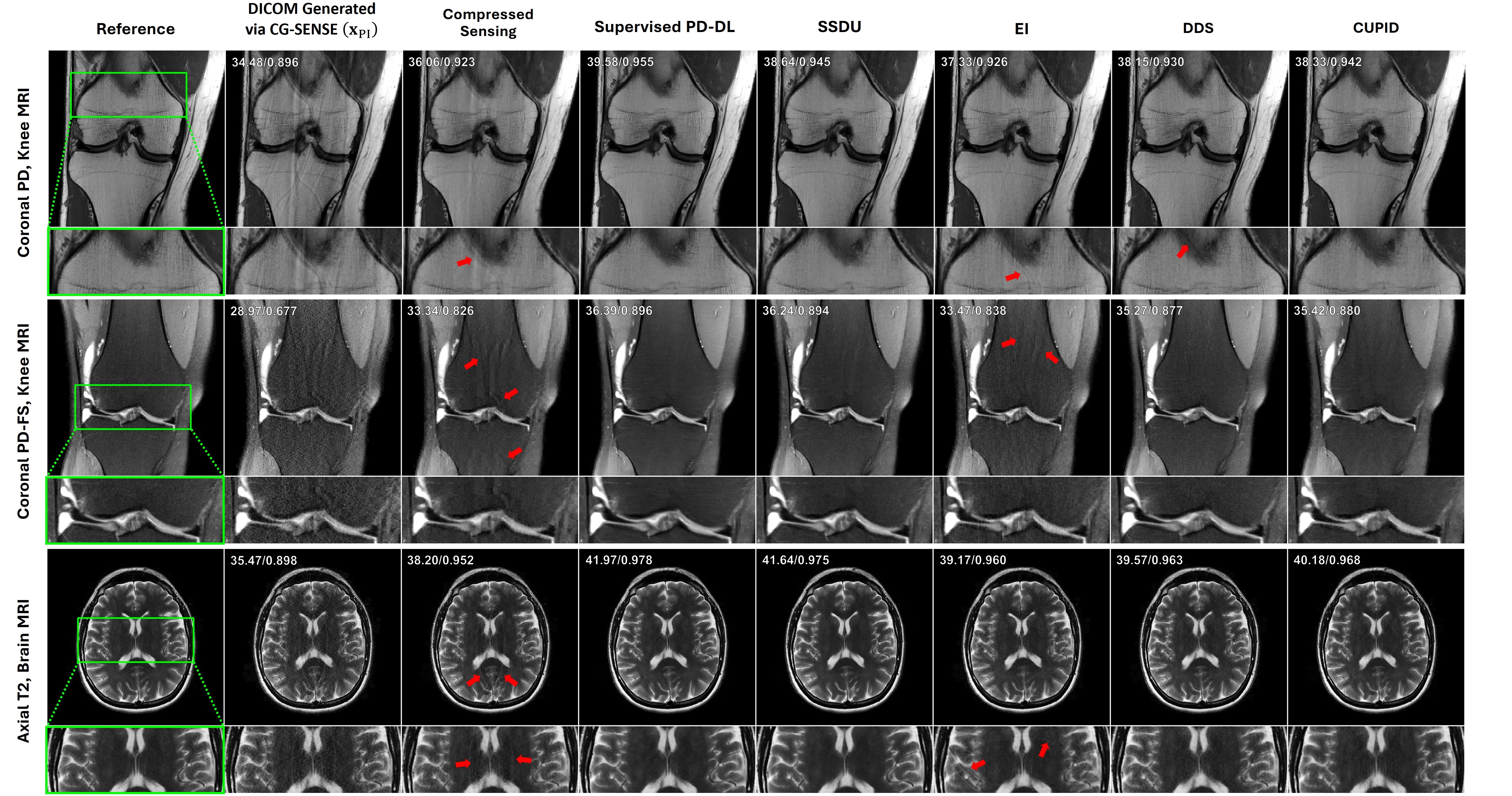}
    \caption{Representative slices reconstructed at an acceleration factor of $R = 4$ using equidistant undersampling from coronal PD and coronal PD-FS knee MRI, as well as axial T2-weighted brain MRI. The baseline CG-SENSE, CS, EI-trained PD-DL, and DDS suffer from residual artifacts highlighted by red arrows. PD-DL trained with CUPID improves upon them while delivering reconstruction quality comparable to supervised and SSDU-trained PD-DL.}
   \label{fig:retrospective}
\end{figure}

In the zero-shot setup, we compared our zero-shot results with zero-shot SSDU (ZS-SSDU)~\cite{yaman2022zeroshot} as well as CS, and our baseline method, CG-SENSE - all of which are compatible with zero-shot inference. We also include DDS and ScoreMRI for inference in this setup, while noting that these diffusion priors have been pre-trained on very large databases. This is done to highlight CUPID's performance and versatility, which is trained in a zero-shot manner on the given slice. All quantitative evaluations used structural similarity index (SSIM) and peak signal-to-noise ratio (PSNR). We further emphasize that CUPID does not use raw k-space data, and as a result, it does not benefit from the redundancies across multiple coils due to the lack of access to multi-coil raw k-space data. Therefore, resources are further restricted, unlike the comparison methods.

\begin{table}[t]
    \caption{Quantitative results for comparison methods on Coronal PD, Coronal PD-FS, Ax FLAIR, and Ax T2 datasets using equispaced undersampling pattern at $R=4$. \underline{First 3 rows}: Gold standard setup. Trained on the same organ, contrast and acceleration rate as the target population using fully-sampled or sub-sampled raw data. \underline{Mid 2 rows:} Trained on the same organ as the target population, using either fully-sampled raw data or fully-sampled artifact-free DICOM images. \underline{Last 3 rows:} No access to raw data; only sub-sampled DICOM images with artifacts. The \colorbox{best}{{\bf best}} and \colorbox{2ndbest}{second-best} values are highlighted, excluding the gold standard setup.}
    \vspace{0.3em}
    \label{tab:quantitative}
    \centering
    \footnotesize
    \setlength{\tabcolsep}{4.1pt}
    \begin{tabular}{@{}lcccccccc@{}}
        \arrayrulecolor{black} \toprule
        \multirow{2}{*}{Method} & \multicolumn{2}{c}{Cor PD, Knee MRI} & \multicolumn{2}{c}{Cor PD-FS, Knee MRI} & \multicolumn{2}{c}{Ax FLAIR, Brain MRI} & \multicolumn{2}{c}{Ax T2, Brain MRI}\\
        \cmidrule(lr){2-3} \cmidrule(lr){4-5} \cmidrule(lr){6-7} \cmidrule(lr){8-9}
        & PSNR$\uparrow$ & SSIM$\uparrow$ & PSNR$\uparrow$ & SSIM$\uparrow$ & PSNR$\uparrow$ & SSIM$\uparrow$ & PSNR$\uparrow$ & SSIM$\uparrow$ \\\midrule
        Supervised~\cite{hammernik2018VarNet} & 40.44 & 0.964 & 35.72 & 0.893 & 37.91 & 0.967 & 36.77 & 0.933 \\
        SSDU~\cite{yaman2020SSDU} & 39.64 & 0.957 & 35.68 & 0.892 & 37.55 & 0.964 & 36.59 & 0.931 \\
        EI~\cite{chen2021equivariant} & 38.07 & 0.938 & 33.83 & 0.849 & 36.40 & 0.953 & 34.58 & 0.909 \\
        \addlinespace[2pt]
        \arrayrulecolor{gray} \hdashline
        \addlinespace[2pt]
        ScoreMRI~\cite{chung2022scoreMRI} & 37.85 & 0.928 & \secondbest{35.01} & \best{0.883} & 35.26 & 0.934 & 33.83 & 0.899 \\
        DDS~\cite{chung2024decomposed} & \secondbest{38.64} & \secondbest{0.950} & 34.89 & 0.875 & \secondbest{36.18} & \secondbest{0.950} & \secondbest{35.16} & \secondbest{0.914} \\
        \arrayrulecolor{gray} \midrule
        PI~\cite{pruessmann2001cgsense,griswold2002grappa} & 35.51 & 0.909 & 29.48 & 0.735 & 31.97 & 0.906 & 31.02 & 0.833 \\
        CS~\cite{lustig2007sparse} & 36.92 & 0.922 & 33.31 & 0.841 & 33.17 & 0.929 & 33.61 & 0.897 \\
        CUPID \textbf{(ours)} & \best{38.82} & \best{0.952} & \best{35.04} & \secondbest{0.880} & \best{36.49} & \best{0.957} & \best{35.31} & \best{0.921} \\
        \arrayrulecolor{black} \bottomrule
    \end{tabular}
\end{table}

\subsection{Experiments with retrospective undersampling}
\paragraph{Database results} Representative results in \figref{fig:retrospective} show that baseline CG-SENSE, CS and EI reconstructions exhibit residual artifacts, whereas DDS suffers from minor noise amplification. In contrast, CUPID successfully eliminates these artifacts from the CG-SENSE image using a well-trained PD-DL network, achieving a state-of-the-art reconstruction quality comparable to supervised PD-DL and SSDU, despite only having access to ${\bf x}_\textrm{PI}$ for training, and not to raw k-space data unlike these other methods. We observe that parallel imaging reconstruction is not clinically usable at higher acceleration rates, but is improved using a CUPID-trained PD-DL reconstruction. Quantitative results in \tabref{tab:quantitative} support visual observations, showing that CUPID consistently outperforms CG-SENSE, CS, and EI across multiple datasets. It also outperforms ScoreMRI and DDS across various scenarios in the majority of cases, while achieving performance comparable to supervised PD-DL and SSDU, both of which have access to raw data. Further qualitative results are provided in \appref{appx:more_qualitative}.

\begin{figure}[!b]
    \centering
    \includegraphics[width=0.99\textwidth]{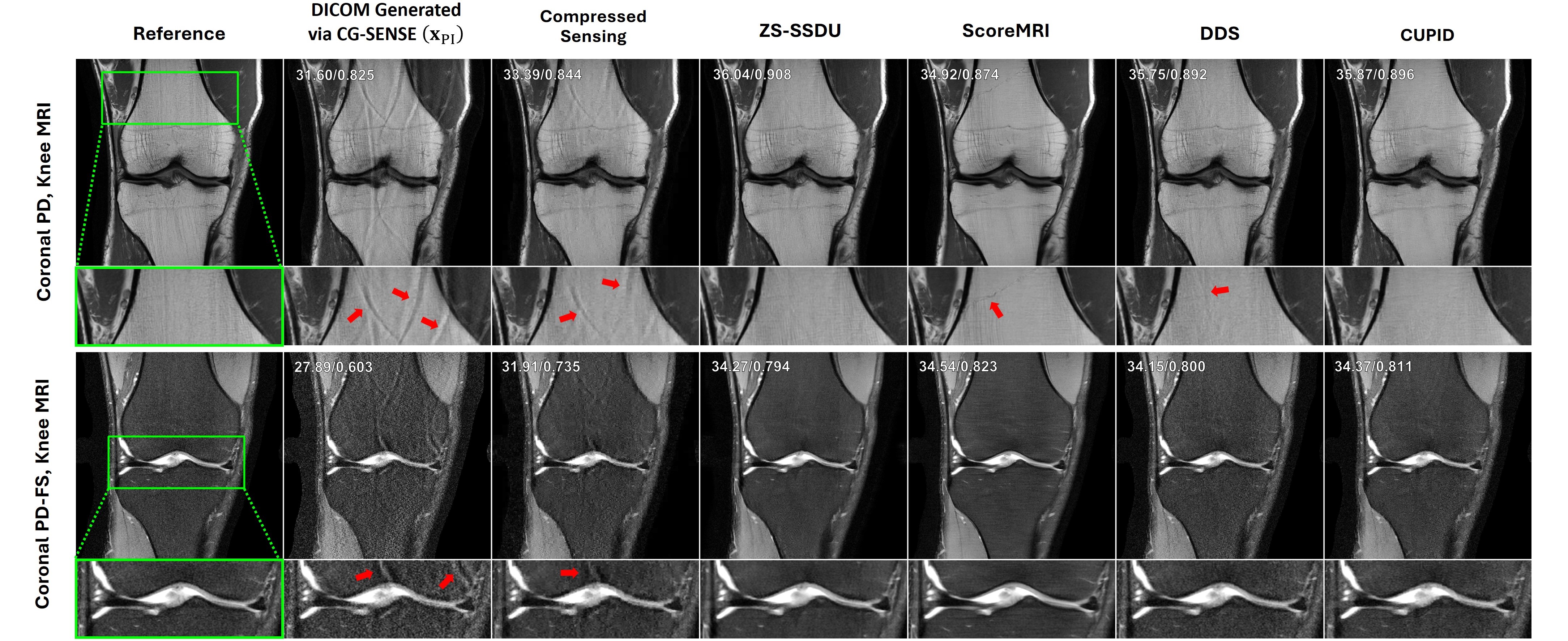}
    \caption{Representative subject-specific/zero-shot learning results for various algorithms on coronal PD and coronal PD-FS knee MRI for retrospective $R=4$ equidistant undersampling, along with results from database-trained diffusion models on knee data. Baseline CG-SENSE, CS, ScoreMRI and DDS suffer from residual artifacts (red arrows) and blurring. PD-DL with CUPID loss successfully removes these artifacts, and functions in a similar manner to ZS-SSDU.}
   \label{fig:zero-shot}
\end{figure}

\paragraph{Subject-specific/zero-shot learning results} \figref{fig:zero-shot} shows zero-shot reconstruction results. CG-SENSE and CS suffer from noise amplification and residual artifacts, with CG-SENSE showing more severe degradation. ScoreMRI and DDS exhibit artifacts in coronal PD, and ScoreMRI introduces blurring in coronal PD-FS. This blurring leads to higher distortion metrics (\eg, PSNR/SSIM), which favor smoother outputs due to penalties on high-frequency details, reflecting the perception-distortion trade-off~\cite{blau2018perception,chung2023dps,alcalar2024ZAPS}. CUPID again achieves superior artifact and noise reduction, closely matching ZS-SSDU quality despite lacking raw data or self-validation.

\paragraph{Ablation studies}
We further conducted three ablation studies to examine key factors influencing CUPID’s performance.
The first examined the effect of the number of perturbations (\appref{appx:perturbations}), the second explored the impact of the hyperparameter $\lambda$ across multiple values (\appref{appx:lambda_effect}), and the third evaluated CUPID’s robustness to different sampling patterns and higher acceleration factors (\appref{appx:high_R}).

\subsection{Practical setting: Prospective undersampling} \label{sec:prospective}
\begin{wrapfigure}[22]{r}{0.47\textwidth}
\begin{minipage}{0.47\textwidth}
    \centering
    \vspace{-2.3em}
    \includegraphics[width=0.92\linewidth]{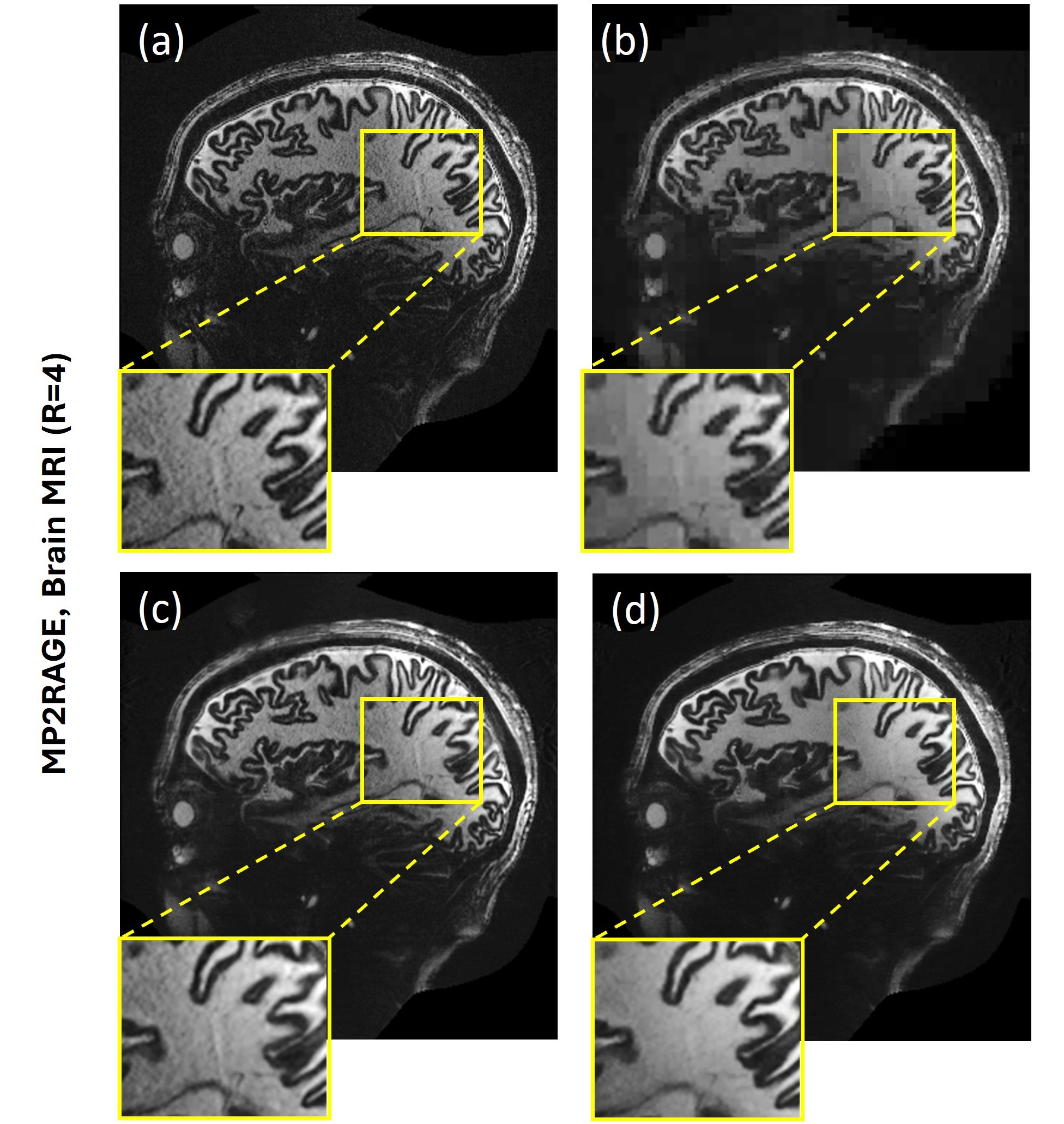}
    \caption{Prospective acceleration results for various methods that operate on parallel imaging-reconstructed DICOMs exported from the scanner. (a) Vendor-provided PI DICOM ($R=4$), (b) CS, (c) DDS, and (d) CUPID.}
   \label{fig:prospective}
\end{minipage}
\end{wrapfigure}

As discussed in \secref{sec:exp_setup}, brain data is acquired at the target acceleration rate, reconstructed via parallel imaging and exported in DICOM format for zero-shot fine-tuning. \figref{fig:prospective} shows reconstruction results for the vendor parallel imaging reconstruction, as well as CS, DDS and CUPID. Vendor-provided $R=4$ reconstruction on the scanner exhibits residual artifacts in \figref{fig:prospective}a (shown in zoomed insets). CS effectively reduces noise but results in an overly-smooth reconstruction. DDS slightly mitigates the artifact and reduces noise, but both remain present. Our proposed CUPID method delivers the most effective artifact and noise mitigation in the DICOM image without requiring any raw k-space data, demonstrating its real-world effectiveness. We note that ZS-SSDU cannot be applied here due to the unavailability of raw data. We also note that the vendor-provided DICOM was generated using k-space interpolation~\cite{griswold2002grappa} instead of the image domain formulation in \eqnref{eq:leastsq_PI}. Due to their equivalence, this did not cause any issues for CUPID, as expected. Furthermore, vendor-provided DICOMs may also include additional zero-padding \cite{shimron2022implicit}, which our physics-based approach handles naturally, as well as additional filtering/processing. More information related on these aspects is given in \appref{appx:grappa} and \appref{appx:limitations}.

\subsection{Out-of-distribution (OOD) example}
\begin{wrapfigure}[15]{r}{0.45\textwidth}
\begin{minipage}{0.45\textwidth}
    \centering
    \vspace{-4.2em}
    \includegraphics[width=0.932\linewidth]{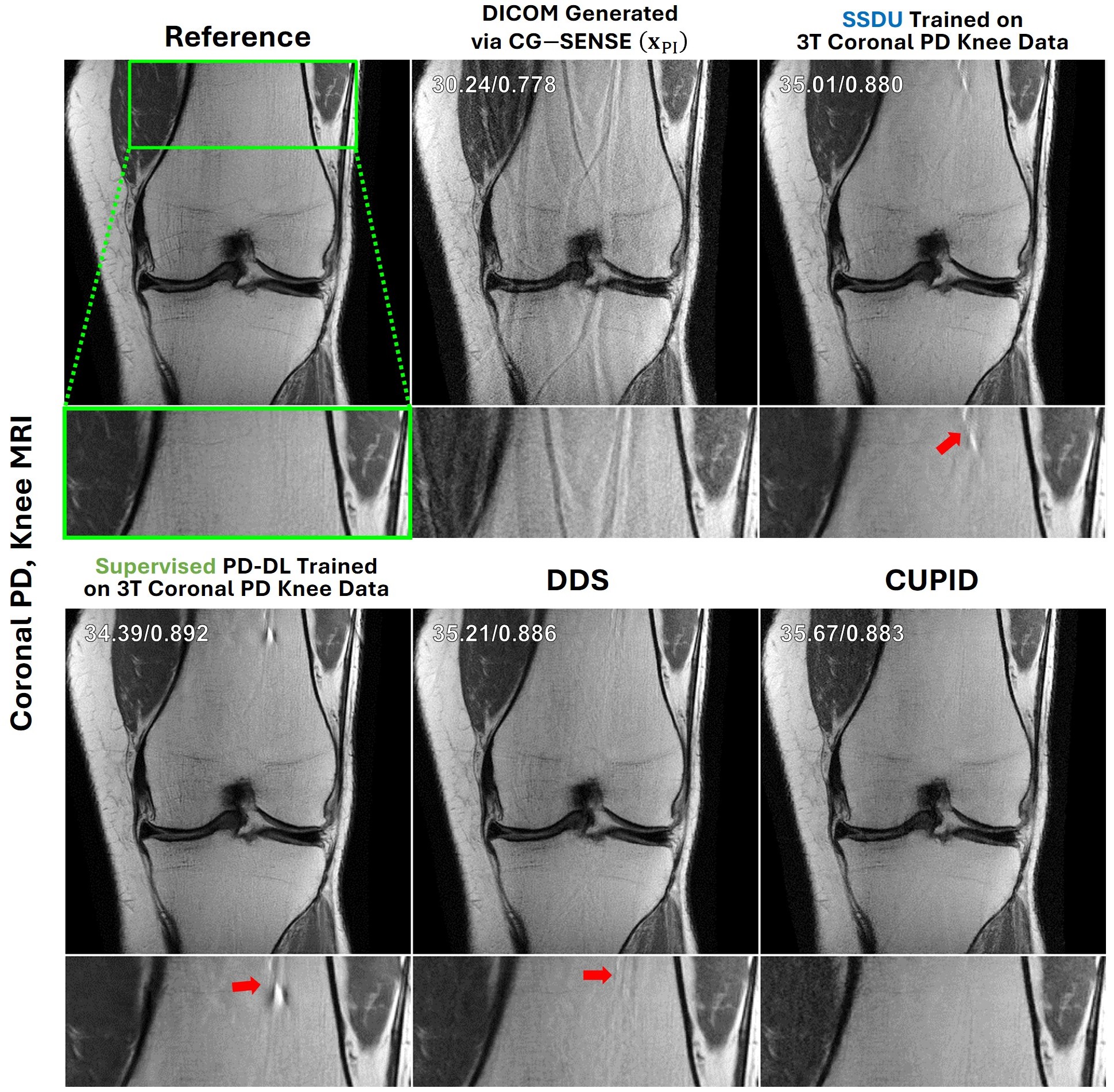}
    \caption{1.5T OOD example showing CUPID's superior artifact reduction.}
    \label{fig:1-5T}
\end{minipage}
\end{wrapfigure}
We further illustrate the generalizability issues discussed in~\secref{sec:why_CUPID} with a representative out-of-distribution example, intended to emulate the transfer from urban training datasets to inference in a rural setting, where scanner hardware and acquisition protocols often differ. Specifically, in \figref{fig:1-5T}, we demonstrate how SSDU and supervised PD-DL models trained on 3T coronal PD data struggle when applied to 1.5T data with a different SNR, leading to noticeable artifacts. DDS, trained on both 3T and 1.5T data, does a better job of mitigating this issue, though some performance degradation persists. CUPID, zero-shot trained on the given slice without the need for raw data, performs the best, effectively reducing artifacts and maintaining high image quality.

\begin{figure}[!t]
    \centering
    \includegraphics[width=1.0\textwidth]{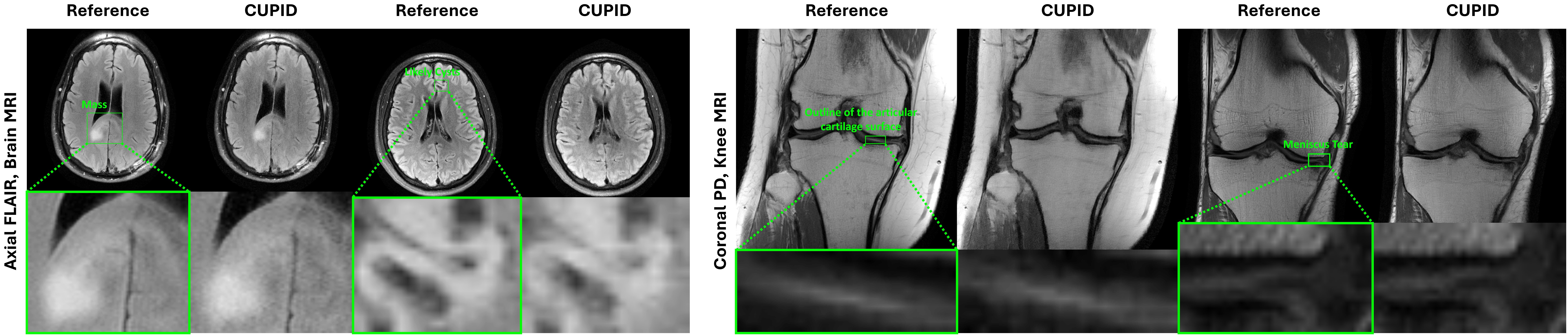}
    \caption{Representative pathology cases from the FastMRI+ dataset~\cite{zhao2022fastmri_plus} showing CUPID reconstructions alongside reference images for brain and knee pathologies. Qualitative expert radiologist assessments suggest that CUPID maintains diagnostic content and visual fidelity across pathology regions of interest.}
   \label{fig:pathology}
\end{figure}

\subsection{Qualitative radiologist readings including FastMRI+ pathology cases}
To assess the clinical plausibility of CUPID’s results, we consulted a musculoskeletal radiologist and a neuroradiologist, whose areas of expertise align with the knee and brain datasets. Both experts independently and blindly reviewed the reconstructed images presented in this work, as well as representative annotated pathology cases from the FastMRI+ dataset~\cite{zhao2022fastmri_plus}, and provided an evaluation of diagnostically relevant contents and overall clinical realism.

In retrospective experiments regarding database results (\figref{fig:retrospective}), the musculoskeletal radiologist observed that CUPID led to slightly more blurring than the supervised and self-supervised PD-DL methods, though this was not deemed diagnostically significant. Furthermore, CUPID was judged superior to DDS for noise suppression and artifact reduction by both radiologists. For the zero-shot experiments (\figref{fig:zero-shot}), the musculoskeletal radiologist reported no noticeable blurring relative to ZS-SSDU and again found CUPID to yield lower noise than DDS.

For the prospective reconstructions (\figref{fig:prospective}), reconstructions (C) and (D) were favored for their lower noise levels, whereas (A) was identified as the noisiest. The neuroradiologist preferred (D) overall, citing clearer gray–white matter boundaries and less pixelation compared to the other variants.

Finally, in the pathology evaluation (\figref{fig:pathology}), representative FastMRI+ cases were reviewed by the same sub-specialists. For the brain dataset, the neuroradiologist compared CUPID and reference reconstructions of a mass and a likely cyst, noting no visible differences in the pathology regions and confirming that diagnostic information was maintained. For the knee dataset, the musculoskeletal radiologist assessed meniscus tear (complex degenerative tearing of the body of the meniscus with meniscal extrusion) and detailed depiction of the articular surface, which is critical for evaluating cartilage integrity. In both cases, CUPID reconstructions were considered diagnostically similar to the references in the clinically relevant areas.

\section{Conclusion}
In this study, we proposed Compressibility-inspired Unsupervised Learning via Parallel Imaging Fidelity (CUPID), a novel training strategy for PD-DL MRI reconstruction using only clinically accessible images, without raw k-space data. CUPID leverages image compressibility and carefully designed perturbations that preserve parallel imaging consistency, enabling high-quality, physics-driven reconstructions. To our knowledge, this is the first method to train PD-DL networks solely from clinical images. CUPID also alleviates the training burden of generative methods, which requires a large number of data during training to capture the prior well. Extensive retrospective and prospective evaluations demonstrate the effectiveness of CUPID across diverse MRI scans and learning settings.

\paragraph{Acknowledgements}
The authors gratefully acknowledge Dr. Jutta Ellermann and Dr. Can \"{O}z\"{u}temiz for providing expert radiologist assessments during the rebuttal stage on short notice. 

\bibliographystyle{abbrv}
\small
\bibliography{refs}
\newpage
\normalsize
\appendix
\section{Implementation Details for Each Method} \label{appx:imp_details}
\paragraph{PD-DL based approaches} For each PD-DL method, VS-QP \cite{aggarwal2019MoDL,yaman2020SSDU,demirel2021EMBC_20fold_7TfMRI,demirel2023SIIM,yaman2021_3D-LGE} was unrolled to solve \eqnref{eq:least_sq_mri}, transforming it into 2 sub-problems:
\begin{subequations}
    \begin{equation}
        \mathbf{z}^{(i)} = \arg\min_{\bf{z}} \|\mathbf{x}^{(i-1)} -\mathbf{z}\|_2^2 + \mathcal{R}(\mathbf{z}), \label{eq:vsqp_1}
    \end{equation}
    \begin{equation}
        \mathbf{x}^{(i)} = \arg\min_{\bf x} \|\mathbf{y}_\Omega - \mathbf{E}_\Omega \mathbf{x}\|_2^2 + \mu \|\mathbf{x} - \mathbf{z}^{(i)}\|_2^2, \label{eq:vsqp_2}
    \end{equation}
\end{subequations}
where \eqnref{eq:vsqp_1} is the proximal operator for the regularization, implicitly solved using neural networks, while \eqnref{eq:vsqp_2} accounts for data fidelity and has a closed form solution:
\begin{equation}
    \mathbf{x}^{(i)} = \left( \mathbf{E}^H_{\Omega} \mathbf{E}_{\Omega} + \mu \mathbf{I} \right)^{-1} (\mathbf{E}^H_{\Omega} \mathbf{y}_\Omega + \mu \mathbf{z}^{(i)}), \label{eq:closed_form}
\end{equation}
which was solved using a conjugate-gradient (CG) method ~\cite{aggarwal2019MoDL} with $15$ iterations. Unrolled network for each method comprised $10$ unrolls, while the regularizer was implemented as a CNN-based ResNet architecture~\cite{timofte2017ntire} that had 15 residual blocks. Each layer within these blocks had 3$\times$3 kernels and 64 channels, totaling 592,129 trainable parameters. The unrolled network was trained in an end-to-end fashion for 100 epochs. For supervised PD-DL~\cite{hammernik2018VarNet,aggarwal2019MoDL}, the normalized $\ell_1$-$\ell_2$ loss function was used between the reconstructed and ground truth raw k-space data~\cite{knoll2020deep-survey, demirel2023NER}. For SSDU, $\rho=|\Delta|/|\Omega|=0.4$ was used as proposed in~\cite{yaman2020SSDU}. For EI~\cite{chen2021equivariant}, we modified the loss function in PD-DL networks to:
\begin{equation}
    \min_{\boldsymbol{\theta}} \mathbb{E} \left[ \mathcal{L} \left( \mathbf{y}_{\Omega}, \hat{\mathbf{x}}_\Omega \right) \right] + \beta \sum_{g \in G} \mathcal{L} \left( \mathcal{T}_g \hat{\mathbf{x}}_\Omega, f \left( \mathbf{E}_{\Omega} \mathcal{T}_g \hat{\mathbf{x}}_\Omega, \mathbf{E}_{\Omega}; \boldsymbol{\theta} \right) \right), \label{eq:ei}
\end{equation}
where $\hat{\mathbf{x}}_\Omega = f\left( \mathbf{y}_{\Omega}, \mathbf{E}_{\Omega}; \boldsymbol{\theta} \right)$ is the PD-DL network output. First term in \eqnref{eq:ei} enforces consistency with acquired raw data, while the second term imposes equivariance relative to a group of transformations, $\{\mathcal{T}_g\}_{g \in G}$. Here, $|G|$ is the cardinality of $\{\mathcal{T}_g\}_{g \in G}$ and $\beta$ is the equivariance weight. We followed the authors' publicly available CT reconstruction code for EI~\cite{chen2021equivariant}, and employed 3 rotations along with 2 flips. For CUPID, dual-tree complex wavelet transform (DTCWT), which provides an over-complete representation~\cite{selesnick2005dualtree-WT, cotter2020uses}, was selected as the sparsifying transform ($\bf{W}$) in \eqnref{eq:loss1}. Furthermore, $\mathbf{x}^{(0)}$ in \eqnref{eq:loss1}, i.e. the initial estimate prior to any reweighting, was calculated using a CS approach as mentioned in \secref{sec:methods}. This was implemented using \eqnref{eq:cs} with 10 iterations, with 10 CG steps for data fidelity and $0.01 \cdot ||{\bf Wx}_\textrm{PI}||_{\infty}$ as the soft thresholding parameter.
\paragraph{Compressed sensing} We solved the regularized $\ell_1$ minimization problem given below:
\begin{equation}
    \arg\min_{\bf x} \|\mathbf{y}_{\Omega}-\mathbf{E}_{\Omega}\mathbf{x}\|_2 + \tau \|\mathbf{Wx}\|_1, \label{eq:cs}
\end{equation}
using VS-QP~\cite{fessler2020SPM}. Similar to the unrolled network, data fidelity was solved using CG, and soft thresholding was implemented on the DTCWT coefficients.
\paragraph{DDS} We followed the original code and pre-trained score network provided by~\cite{chung2024decomposed} in their corresponding public repository. We used 50-150 DDIM sampling steps depending on the SNR and noise level, and tuned the number of CG iterations for the best performance.
\paragraph{ScoreMRI} For ScoreMRI implementation, we followed the original code and pre-trained network provided by~\cite{chung2022scoreMRI} in their corresponding public repository and used 2000 predictor-corrector (PC) sampling.

Each method (including CUPID) was implemented using a single NVIDIA A100-SXM4-80GB GPU that has a total memory of 80 GB.

\begin{table}[t]
    \caption{Training and inference/run-time of comparison methods under database and zero-shot settings using a single NVIDIA A100-SXM4-80GB GPU. N/A indicates methods without dataset-level training or zero-shot configuration.}
    \vspace{0.3em}
    \label{tab:time}
    \centering
    \footnotesize
    \setlength{\tabcolsep}{6.5pt}
    \begin{tabular}{@{}lccc c@{}}
        \toprule
        \multirow{2}{*}{Method} & \multicolumn{2}{c}{\textbf{Database setting}} & \multicolumn{1}{c}{\textbf{Zero-shot setting}} \\
        \cmidrule(lr){2-3} \cmidrule(lr){4-4}
        & Train [GPU hours] $\downarrow$  & Inference $\downarrow$ & Runtime $\downarrow$ \\ 
        \midrule
        Supervised~\cite{hammernik2018VarNet} 
            & {\color{gray!150} $\sim$3.5 hours} 
            & {\color{gray!150} $\ll1$ seconds} 
            & - \\
        SSDU~\cite{yaman2020SSDU} 
            & {\color{gray!150} $\sim$3.5 hours} 
            & {\color{gray!150} $\ll1$ seconds} 
            & - \\
        EI~\cite{chen2021equivariant} 
            & {\color{gray!150} $\sim$3.5 hours}
            & {\color{gray!150} $\ll1$ seconds} 
            & - \\
        \addlinespace[2pt]
        \arrayrulecolor{gray!150}\hdashline
        \addlinespace[2pt]
        ScoreMRI~\cite{chung2022scoreMRI} 
            & {\color{gray!150} $\sim$168 hours}
            & {\color{gray!150} $\sim$350 seconds} 
            & - \\
        DDS (150 NFEs)~\cite{chung2024decomposed} 
            & {\color{gray!150} $\sim$168 hours} 
            & {\color{gray!150} $\sim$10 seconds} 
            & - \\
        \addlinespace[2pt]
        \arrayrulecolor{gray!150}\hdashline
        \addlinespace[2pt]
        ZS-SSDU~\cite{yaman2020SSDU} 
            & - 
            & - 
            & {\color{gray!150} $\sim \mathrm{K_{ZS-SSDU}}\times$15 seconds} \\
        \arrayrulecolor{gray}\midrule
        CUPID \textbf{(ours)} 
            & {\color{gray!150} $\sim (\mathrm{K_{CUPID}}+1)\times$3.5 hours}
            & {\color{gray!150} $\ll1$ seconds}
            & {\color{gray!150} $\sim (\mathrm{K_{CUPID}}+1)\times$63 seconds} \\
        \arrayrulecolor{black}\bottomrule
    \end{tabular}
\end{table}

\section{Computational Efficiency and Runtime Analysis}
As noted in \secref{sec:comparison_methods} and \appref{appx:imp_details}, all PD-DL comparison methods share the same unrolled network architecture. Consequently, they exhibit identical inference times of under one second per slice once they are trained on a database. Diffusion-based methods, such as ScoreMRI and DDS, require substantially longer training and sampling times due to their iterative denoising processes and large model complexity. These methods often involve hundreds to thousands of sampling steps during inference, resulting in inference times that are orders of magnitude higher than PD-DL approaches.

In the zero-shot setting, ZS-SSDU serves as a useful baseline that, like CUPID, adapts to individual scans without dataset-level retraining. However, ZS-SSDU relies on access to raw k-space data and employs $\mathrm{K_{ZS-SSDU}} = 25$ k-space masks in its public implementation. In contrast, CUPID operates entirely on reconstructed DICOM images and eliminates the need for raw data while maintaining comparable zero-shot reconstruction performance.

\begin{wrapfigure}[22]{r}{0.55\textwidth}
\begin{minipage}{0.55\textwidth}
    \centering
    \vspace{-0.6em}
    \includegraphics[width=0.95\linewidth]{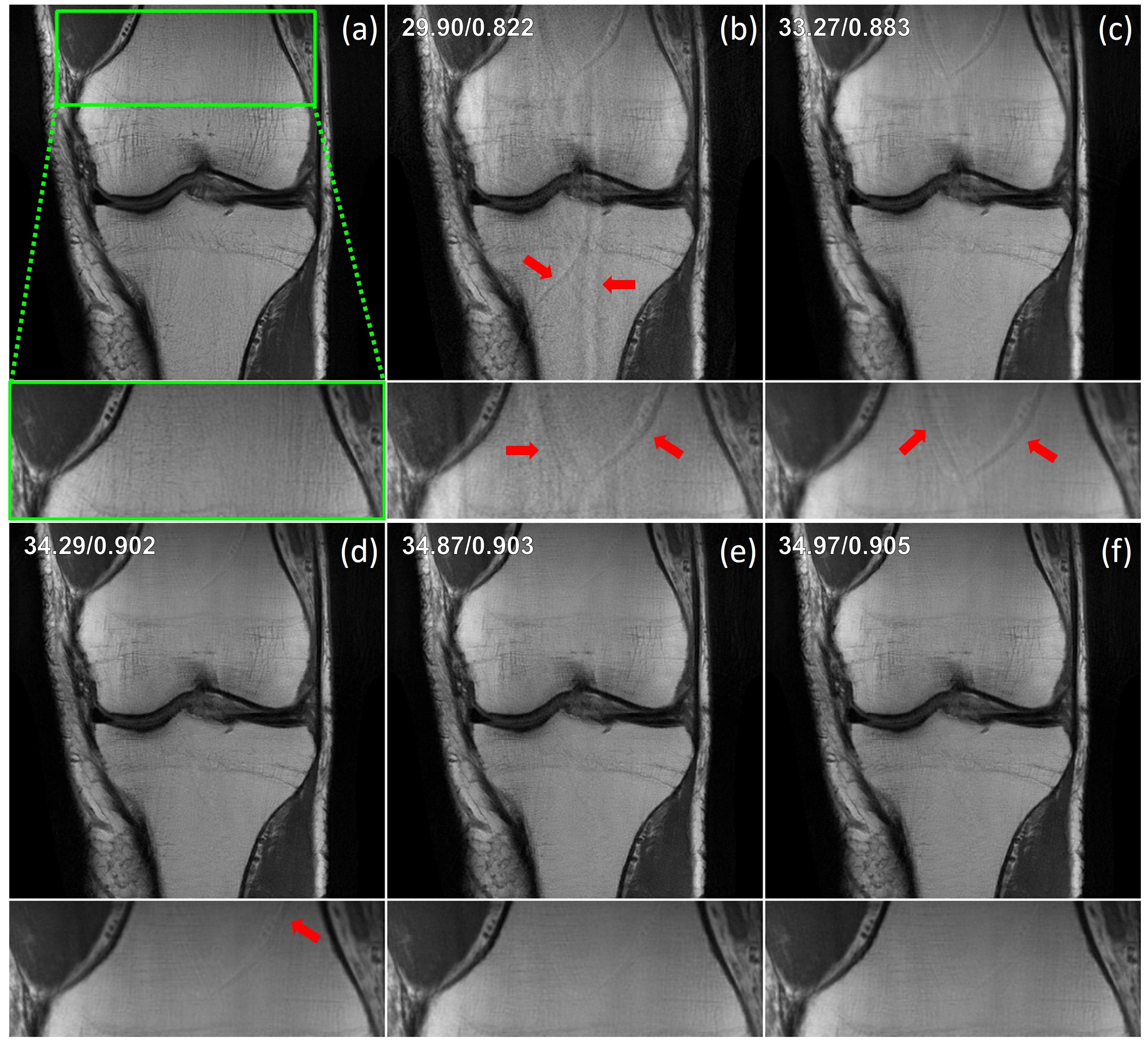}
    \caption{Representative zero-shot fine-tuning result for CUPID with varying perturbation counts ($K$) on coronal PD knee MRI ($R=4$). (a) Fully-sampled, (b) CG-SENSE, (c–f) CUPID with $K=1,3,6,10$. Fewer perturbations ($K<6$) yield artifacts due to poor expectation estimates; quality improves with $K$, but gains become negligible beyond $K=6$.}
    \label{fig:ablation2}
\end{minipage}
\end{wrapfigure}

Although zero-shot optimization introduces additional computation, such run-times are generally acceptable in clinical workflows where reconstructions can be completed offline, for instance when image readings occur on the following day~\cite{basha2017clinical}. Furthermore, CUPID introduces a controllable trade-off between reconstruction quality and computational cost, determined by the number of perturbation patterns used during optimization (denoted as $\mathrm{K_{CUPID}}$ in \tabref{tab:time}). In our experiments, we set $\mathrm{K_{CUPID}} = 6$, resulting in a runtime of approximately 7–8 minutes per slice, depending on the network depth. Both factors can be adjusted to balance reconstruction fidelity against run-time.

\section{Perturbation Strategies}

\subsection{Choice for number of perturbation patterns} \label{appx:perturbations}
\begin{figure}[t]
  \centering
  \begin{subfigure}{0.49\linewidth}
    \includegraphics[width=1.0\textwidth]{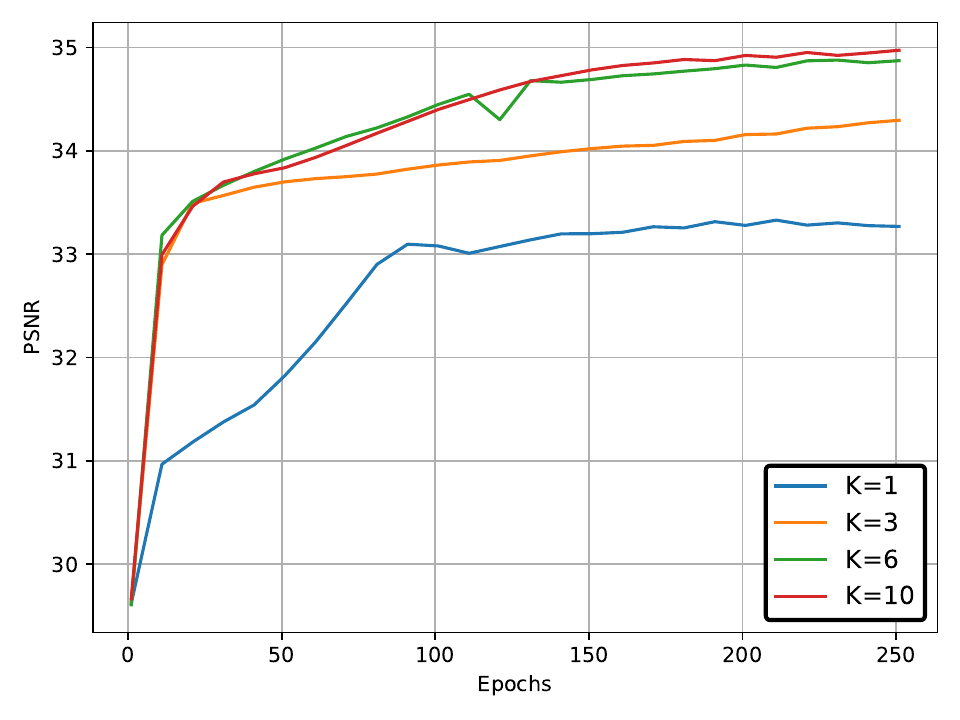}
    \caption{PSNR curves for each $K$ value.}
    \label{fig:ablation2-a}
  \end{subfigure}
  %\hfill
  \begin{subfigure}{0.49\linewidth}
    \includegraphics[width=1.0\textwidth]{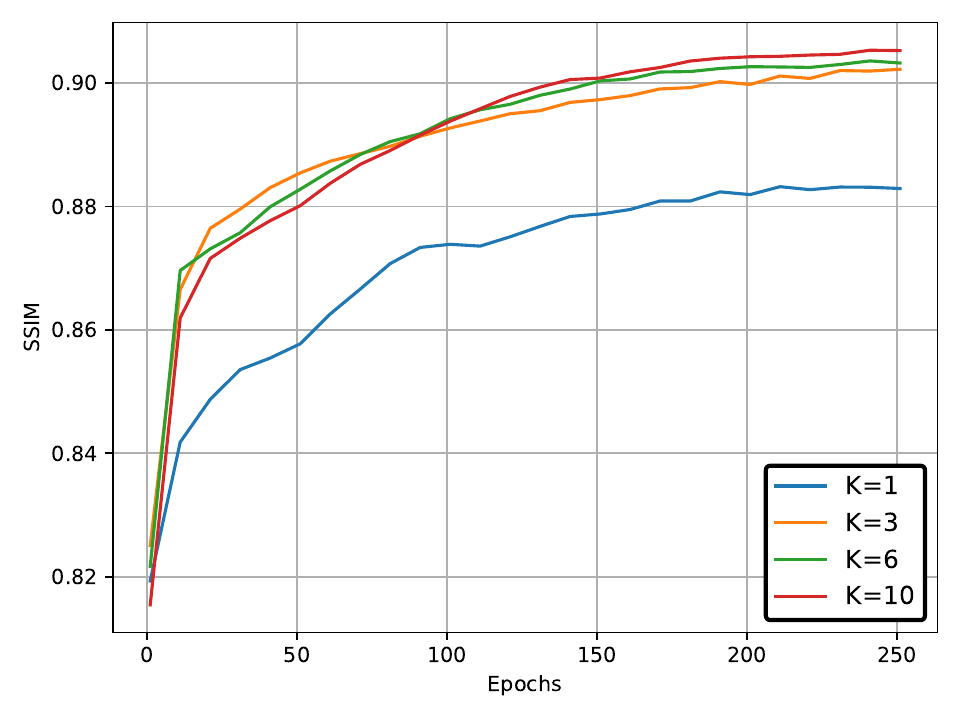}
    \caption{SSIM curves for each $K$ value.}
    \label{fig:ablation2-b}
  \end{subfigure}
  \caption{PSNR and SSIM curves confirm the visual observations with respect to the number of perturbations, $K$. Lower $K$ values tend to perform worse and increasing $K$ becomes redundant after a certain point.}
\end{figure}

The empirical expectation that approximates the one in \eqnref{eq:loss2} is expected to converge to the true expectation as we introduce more perturbation patterns and randomness over the choice of ${\bf p}$. \figref{fig:ablation2} shows the zero-shot fine-tuning results of CUPID with $K \in \{1,3,6,10\}$, while \figref{fig:ablation2-a} and \figref{fig:ablation2-b} illustrates the corresponding PSNR and SSIM curves throughout the training epochs, respectively. As expected, using a single pattern does not capture the true mean and exhibits artifacts. As we introduce more perturbations, we reduce the artifacts and noise amplification. At a certain point, increasing the number of perturbations becomes counterproductive, yielding only marginal gains while significantly increasing the computation time. Thus, we opted to use 6 distinct $\mathbf{p}_\text{k}$ patterns throughout our study as it offers the optimal trade-off.

\subsection{Design alternatives for perturbations} \label{appx:pert_design}
\begin{figure}[t]
  \centering
   \includegraphics[width=1.0\textwidth]{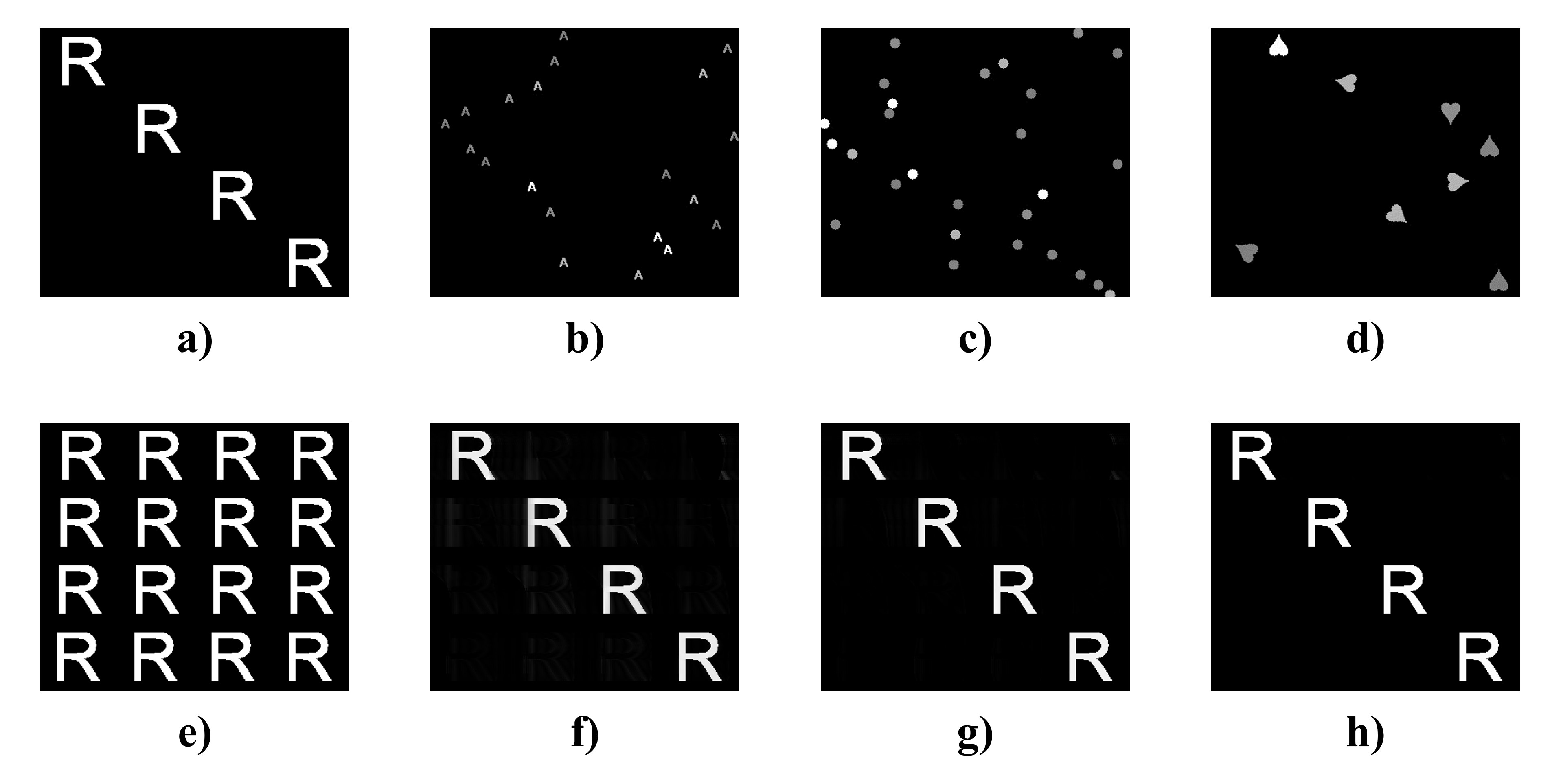}
   \caption{Added perturbations may consist of: (a) precisely positioned letters that have the same intensity in within each shape, (b) randomly positioned letters or (c) circles that have different intensities, or (d) randomly rotated card suits. Furthermore, when accelerated by $R = 4$, even without ACS data, whose corresponding zerofilled image is shown in (e), these perturbations can be resolved via parallel imaging methods such as CG-SENSE: (f) 20 iterations, (g) 40 iterations, (h) 80 iterations.}
   \label{fig:pert_design_alt}
\end{figure}

As stated in \secref{sec:methods}, added perturbations may consist of several different structures. \figref{fig:pert_design_alt} provides some of these perturbation examples, an illustration of how the perturbation looks with undersampling, and how they are recovered perfectly through conventional parallel imaging methods.

We note that there was no task-specific perturbation that we used, meaning that the perturbations selected from the same set were applied to all datasets given that the created perturbations do not create fold-overs at R-fold which result in artifacts. Note the latter condition means they should be recoverable through parallel imaging reconstruction. Finally, we note that when calculating the sample mean estimate for \eqnref{eq:loss2}, intensity of the perturbations was empirically found to be more important than their shapes/orientations. Specifically, we observed that varying it randomly within the perturbation, as in \figref{fig:pert_design_alt}b-d, leads to improved reconstruction outcomes.

\section{Compatibility with Various Parallel Imaging Reconstructions} \label{appx:grappa}
Vendor reconstructions typically use different parallel imaging techniques. For our retrospective studies, we used CG-SENSE (or equivalently SENSE)~\cite{pruessmann1999sense} because it naturally fits with the DF units in the unrolled network, and it is commonly used in clinical settings, alongside GRAPPA~\cite{griswold2002grappa}. However, we emphasize that our method does not make assumptions about the specific reconstruction method used by the vendor; instead, it assumes that parallel imaging can resolve the perturbations, which is ensured by designing them in a manner that prevents fold-over aliasing artifacts from overlapping.

\begin{figure}[t]
  \centering
   \includegraphics[width=0.8\linewidth]{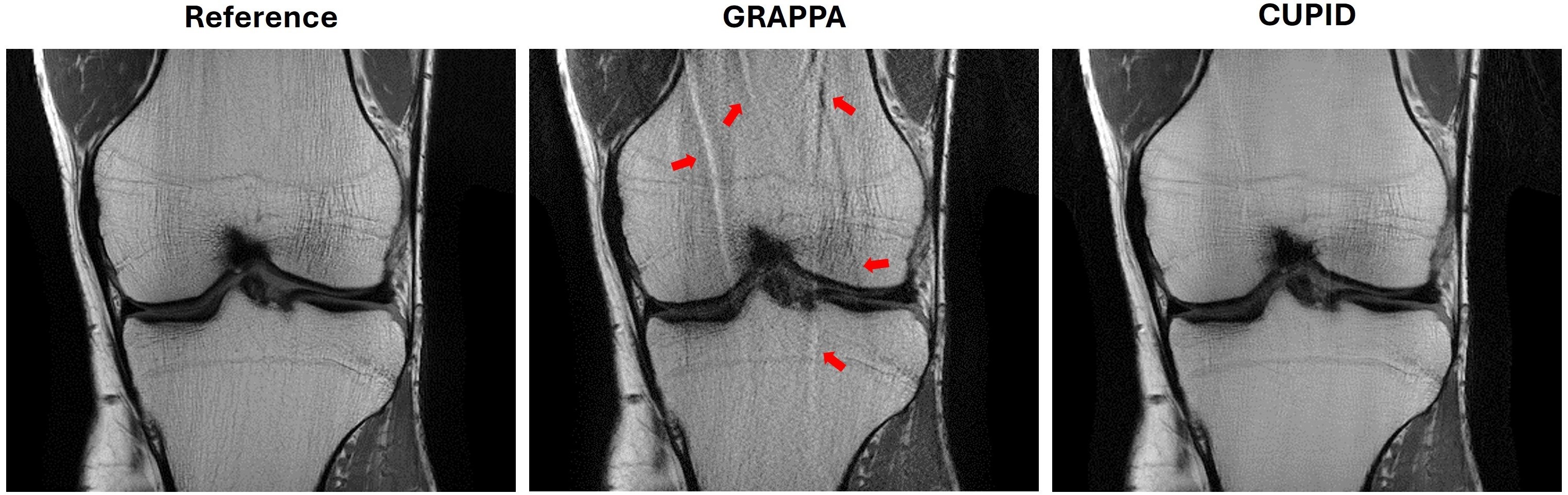}
   \caption{Representative reconstructions for CUPID with $\mathbf{x}_\textrm{PI}$ reconstructed using GRAPPA on coronal PD knee MRI using $R=4$ uniform undersampling. GRAPPA exhibits aliasing and noise artifacts at this high acceleration rate. PD-DL network trained with a CUPID implementation that only has access to this GRAPPA reconstruction improves on it, reducing these artifacts. This highlights the compatibility of CUPID with different parallel imaging reconstructions.}
   \label{fig:grappa}
\end{figure}

To further validate this, we include representative CUPID reconstruction results in \figref{fig:grappa} where $\mathbf{x}_\text{PI}$ is generated via GRAPPA~\cite{griswold2002grappa}, demonstrating that CUPID is compatible with different types of parallel imaging reconstructions as input. We further note that the prospective study also used GRAPPA reconstruction as input, as this is the reconstruction provided by the vendor used in our institution.

\begin{figure}[!b]
    \centering
    \includegraphics[width=\linewidth]{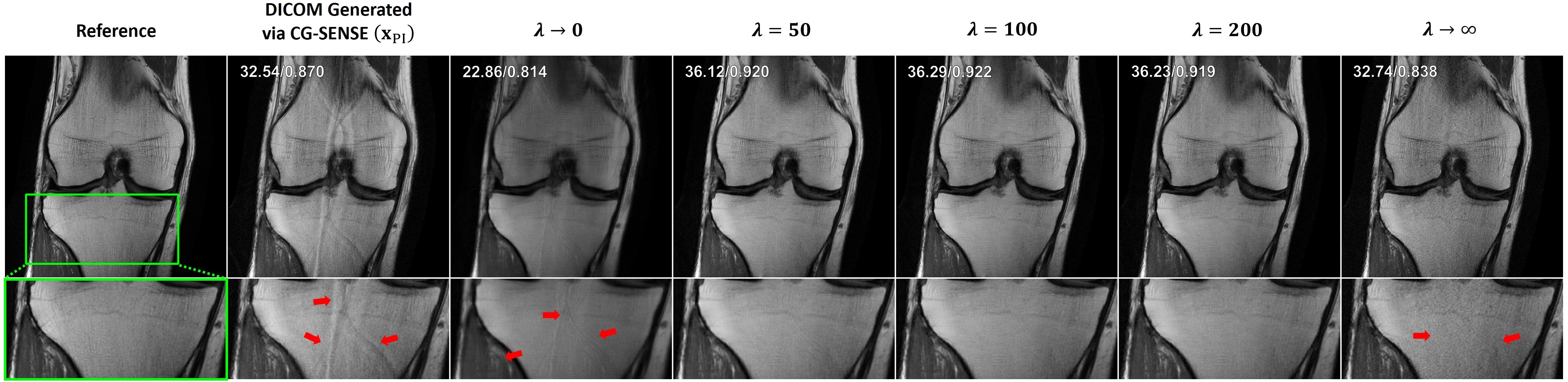}
    \caption{Using only the compressibility term ($\lambda = 0$) in the loss leads to overly-smoothed images, whereas using only the parallel imaging fidelity term ($\lambda \to \infty$) causes noise amplification (red arrows). Using both terms with a mid-range $\lambda$ value as a trade-off provides high-quality reconstructions that are clear from noise and artifacts.}
    \label{fig:ablation1}
\end{figure}

\section{Effect of the Trade-off Parameter} \label{appx:lambda_effect}
We explored the effect of $\lambda$ parameter in CUPID by training 5 distinct PD-DL networks using $\lambda \in \{0,50,100,200,\infty\}$. We note that using $\lambda = 0$ corresponds to using only the compressibility term ($\mathcal{L}_{\text{comp}}$ in \eqnref{eq:loss1}), whereas using $\lambda \to \infty$ translates to using solely the parallel imaging fidelity term ($\mathcal{L}_{\text{pif}}$ in \eqnref{eq:loss2}). \figref{fig:ablation1} shows the corresponding reconstruction results for each case. As outlined in \secref{sec:methods}, only using $\mathcal{L}_{\text{comp}}$ leads to overly-smooth reconstructions due to network forcing the wavelet coefficients towards zero without maintaining consistency with the data. On the other hand, solely using $\mathcal{L}_{\text{pif}}$ results in DIP-like reconstructions~\cite{ulyanov2018_DIP}, where the network overfits the data without any regularization, resulting in noise amplification. CUPID with $\lambda \in \{50,100,200\}$ combines both loss terms to attain high-fidelity reconstructions. Thus, we conclude that CUPID demonstrates robust performance across a wide range of $\lambda$ values, provided that $\lambda$ is chosen within a reasonable range.

%\section{Experiments with Different Acceleration Rates and Patterns}
\section{Experiments on Sampling Pattern Variations and High Acceleration Rates} \label{appx:high_R}
\begin{wrapfigure}[23]{r}{0.66\textwidth}
\begin{minipage}{0.66\textwidth}
    \centering
    \vspace{0.2em}
    \includegraphics[width=0.99\linewidth]{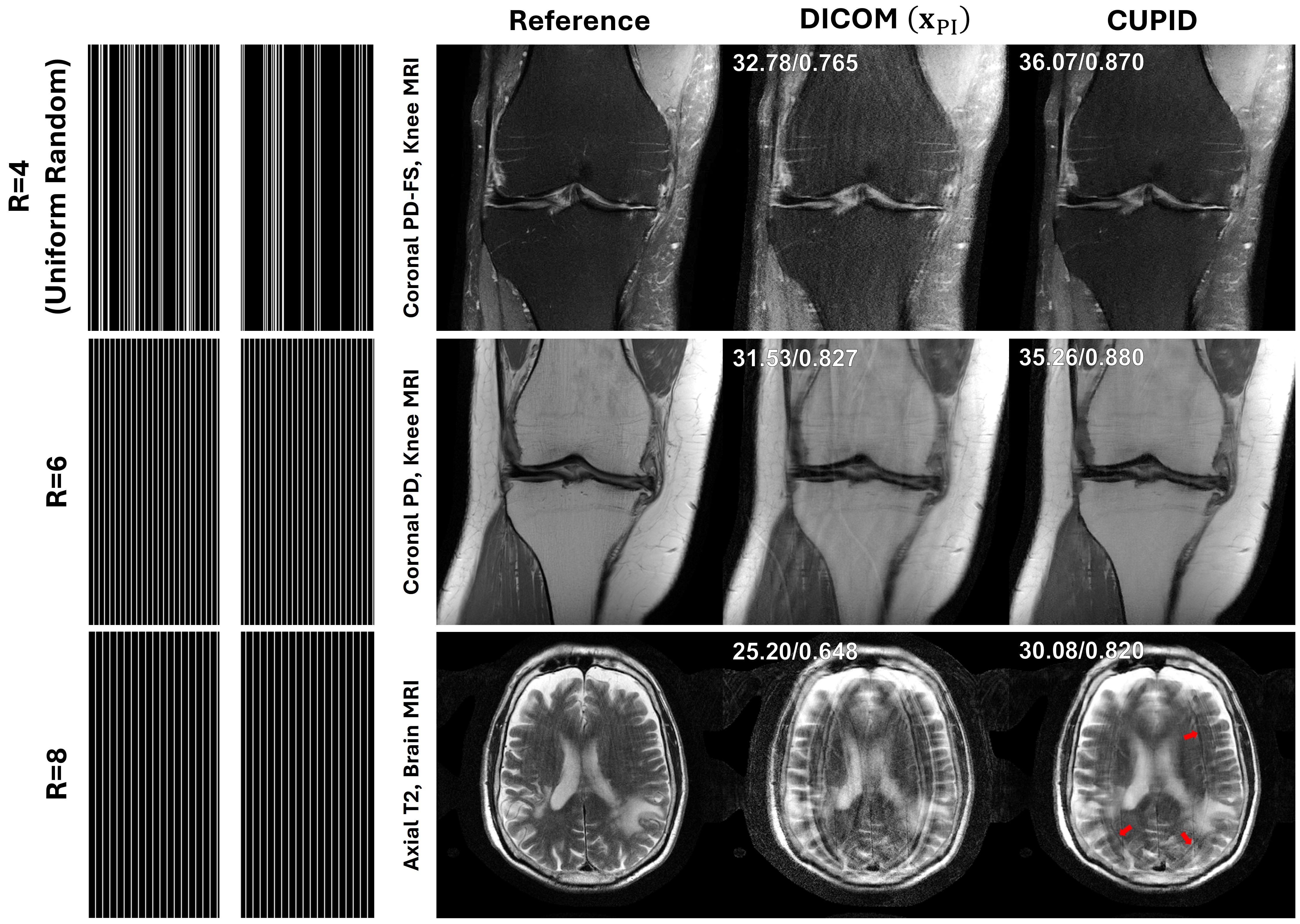}
    \caption{Representative reconstructions from the fastMRI dataset for different acceleration rates and sampling patterns, illustrating robustness up to $R=6$. Artifact emerge at $R=8$, as expected.}
    \label{fig:high_R}
\end{minipage}
\end{wrapfigure}

We further evaluated the robustness of CUPID to variations in acceleration rate and sampling pattern using the fastMRI knee and brain datasets~\cite{knoll2020fastmri_dataset-journal, zbontar2019fastmri_dataset-arXiv}. Representative zero-shot reconstruction results for these settings are shown in Fig.~\ref{fig:high_R}.

Specifically, we tested:
\begin{itemize}
\item Random uniform undersampling at an acceleration rate of $R=4$ (in contrast to the equidistant pattern used in the main text),
\item Equidistant undersampling at higher acceleration rates of $R=6$ and $R=8$.
\end{itemize}

For random uniform undersampling at $R=4$, the results were comparable to those obtained with equidistant sampling, exhibiting similarly clean reconstructions with effective noise suppression and minimal artifacts. At $R=6$, CUPID continued to produce high-quality reconstructions, preserving image sharpness and structure with only minor residual artifacts.

At $R=8$, CUPID still achieved effective noise suppression but exhibited visible residual artifacts in some regions, indicating that this acceleration factor likely represents the practical limit with the current coil configurations and SNRs on the \emph{fastMRI} datasets. The observed degradations arise primarily from the increasing loss of quality for the parallel imaging reconstruction at high acceleration, which weakens the initialization, and from the general limitations of DL-based methods in highly underdetermined regimes. Previous studies have shown that supervised PD-DL methods yield visible blurring and reduced diagnostic quality at $R=8$~\cite{muckley2021fastMRIchallenge-2}, while self-supervised approaches trained with raw k-space data also exhibit residual aliasing at this rate~\cite{yaman2020SSDU, alcalar2025_SPIC_SSDU}. Since CUPID operates without any access to raw k-space measurements, similar or slightly greater degradation at extreme acceleration rates is expected. Overall, these results demonstrate that CUPID generalizes effectively across sampling patterns and maintains strong reconstruction performance up to moderate acceleration factors, with performance degradation at very high acceleration factors consistent with known challenges in highly accelerated MRI reconstruction.

\section{Limitations} \label{appx:limitations}
We note some practical aspects for the translation of CUPID. While our method aims to improve equitable access to fast MRI in low-resource settings, we acknowledge hardware limitations for fine-tuning PD-DL models; however, since CUPID only requires anonymized DICOM images, data can be securely transferred to off-site locations with greater computational resources. In addition, when a reference calibration scan is unavailable, as in most DICOM databases, coil sensitivities must be estimated during reconstruction. This is addressed in standard PD-DL with raw k-space access~\cite{hu2024spicer,arvinte2021deep_j-sense}, but outside the scope of this study.

MR scans may also include filtering operations applied by some vendors that affect the assumption $\mathbf{\hat{x}}_\text{PI} = (\mathbf{E}_\Omega^H\mathbf{E}_\Omega)^{-1}\mathbf{E}_\Omega^H\mathbf{y}_\Omega$. This was discussed extensively in~\cite{shimron2022implicit}, in the context of using retrospective undersampling of DICOM images to train DL reconstruction, especially highlighting the use of zero-padding, which improves the display resolution compared to the acquisition resolution. It was shown that training of models from retrospective undersampling of DICOM image databases for PD-DL training using zero-padding may lead to biases and inaccuracies. Conversely, our approach is physics-driven in nature, and the sampling pattern $\Omega$ naturally accounts for the zero-padding operation. However, our method is not immune to other types of filtering/processing, such as implicit intensity correction~\cite{han2001multi} or deidentification methods~\cite{van2013wu_minn_HCP}, in which case the filtered ${\bf x}_\textrm{PI}$ would need to be treated as the parallel imaging solution corresponding to a filtered version of ${\bf y}$.

\section{Additional Quantitative Analysis with Standart Deviation} \label{appx:std}
To provide a more comprehensive comparison, \tabref{tab:quantitative_with_variances} reports the mean and standard deviation for all quantitative metrics across four datasets. This supplements \tabref{tab:quantitative} by quantifying the consistency of each method.

\begin{table}[!h]
    \caption{Quantitative results with standard deviations for all comparison methods on Coronal PD, Coronal PD-FS, Ax FLAIR, and Ax T2 datasets using equispaced $R=4$ undersampling. Mean $\pm$ std values are shown. Categories and highlight conventions follow \tabref{tab:quantitative} in the main text.}
    \vspace{0.3em}
    \label{tab:quantitative_with_variances}
    \centering
    \tiny
    \setlength{\tabcolsep}{2.1pt}
    \begin{tabular}{@{}lcccccccc@{}}
        \arrayrulecolor{black} \toprule
        \multirow{2}{*}{Method} & \multicolumn{2}{c}{Cor PD, Knee MRI} & \multicolumn{2}{c}{Cor PD-FS, Knee MRI} & \multicolumn{2}{c}{Ax FLAIR, Brain MRI} & \multicolumn{2}{c}{Ax T2, Brain MRI}\\
        \cmidrule(lr){2-3} \cmidrule(lr){4-5} \cmidrule(lr){6-7} \cmidrule(lr){8-9}
        & PSNR$\uparrow$ & SSIM$\uparrow$ & PSNR$\uparrow$ & SSIM$\uparrow$ & PSNR$\uparrow$ & SSIM$\uparrow$ & PSNR$\uparrow$ & SSIM$\uparrow$ \\\midrule
        Supervised~\cite{hammernik2018VarNet} & 40.44 \var{3.27} & 0.964 \var{0.018} & 35.72 \var{2.49} & 0.893 \var{0.058} & 37.91 \var{1.81} & 0.967 \var{0.008} & 36.77 \var{3.25} & 0.933 \var{0.062} \\
        SSDU~\cite{yaman2020SSDU} & 39.64 \var{3.26} & 0.957 \var{0.021} & 35.68 \var{2.49} & 0.892 \var{0.060} & 37.55 \var{1.76} & 0.964 \var{0.009} & 36.59 \var{3.11} & 0.931 \var{0.060} \\
        EI~\cite{chen2021equivariant} & 38.07 \var{3.79} & 0.938 \var{0.033} & 33.83 \var{2.82} & 0.849 \var{0.075} & 36.40 \var{1.70} & 0.953 \var{0.010} & 34.58 \var{2.88} & 0.909 \var{0.066} \\
        \addlinespace[2pt]
        \arrayrulecolor{gray} \hdashline %\midrule
        \addlinespace[2pt]
        ScoreMRI~\cite{chung2022scoreMRI} & 37.85 \var{3.66} & 0.928 \var{0.031} & \secondbest{35.01 \var{2.46}} & \best{0.883 \var{0.058}} & 35.26 \var{1.84} & 0.934 \var{0.013} & 33.83 \var{2.70} & 0.899 \var{0.065} \\
        DDS~\cite{chung2024decomposed} & \secondbest{38.64 \var{3.42}} & \secondbest{0.950 \var{0.026}} & 34.89 \var{2.50} & 0.875 \var{0.064} & \secondbest{36.18 \var{1.95}} & \secondbest{0.950 \var{0.012}} & \secondbest{35.16 \var{3.13}} & \secondbest{0.914 \var{0.076}} \\
        \arrayrulecolor{gray} \midrule
        PI~\cite{pruessmann2001cgsense,griswold2002grappa} & 35.51 \var{4.14} & 0.909 \var{0.047} & 29.48 \var{3.43} & 0.735 \var{0.112} & 31.97 \var{1.99} & 0.906 \var{0.018} & 31.02 \var{2.49} & 0.833 \var{0.075} \\
        CS~\cite{lustig2007sparse} & 36.92 \var{3.72} & 0.922 \var{0.035} & 33.31 \var{2.71} & 0.841 \var{0.079} & 33.17 \var{1.79} & 0.929 \var{0.013} & 33.61 \var{2.61} & 0.897 \var{0.071} \\
        CUPID \textbf{(ours)} & \best{38.82 \var{3.23}} & \best{0.952 \var{0.024}} & \best{35.04 \var{2.45}} & \secondbest{0.880 \var{0.059}} & \best{36.49 \var{1.72}} & \best{0.957 \var{0.009}} & \best{35.31 \var{3.09}} & \best{0.921 \var{0.073}} \\
        \arrayrulecolor{black} \bottomrule
    \end{tabular}
\end{table}

\section{More Results from the Retrospective Study} \label{appx:more_qualitative}
\begin{figure}
    \centering
    \includegraphics[width=\textwidth]{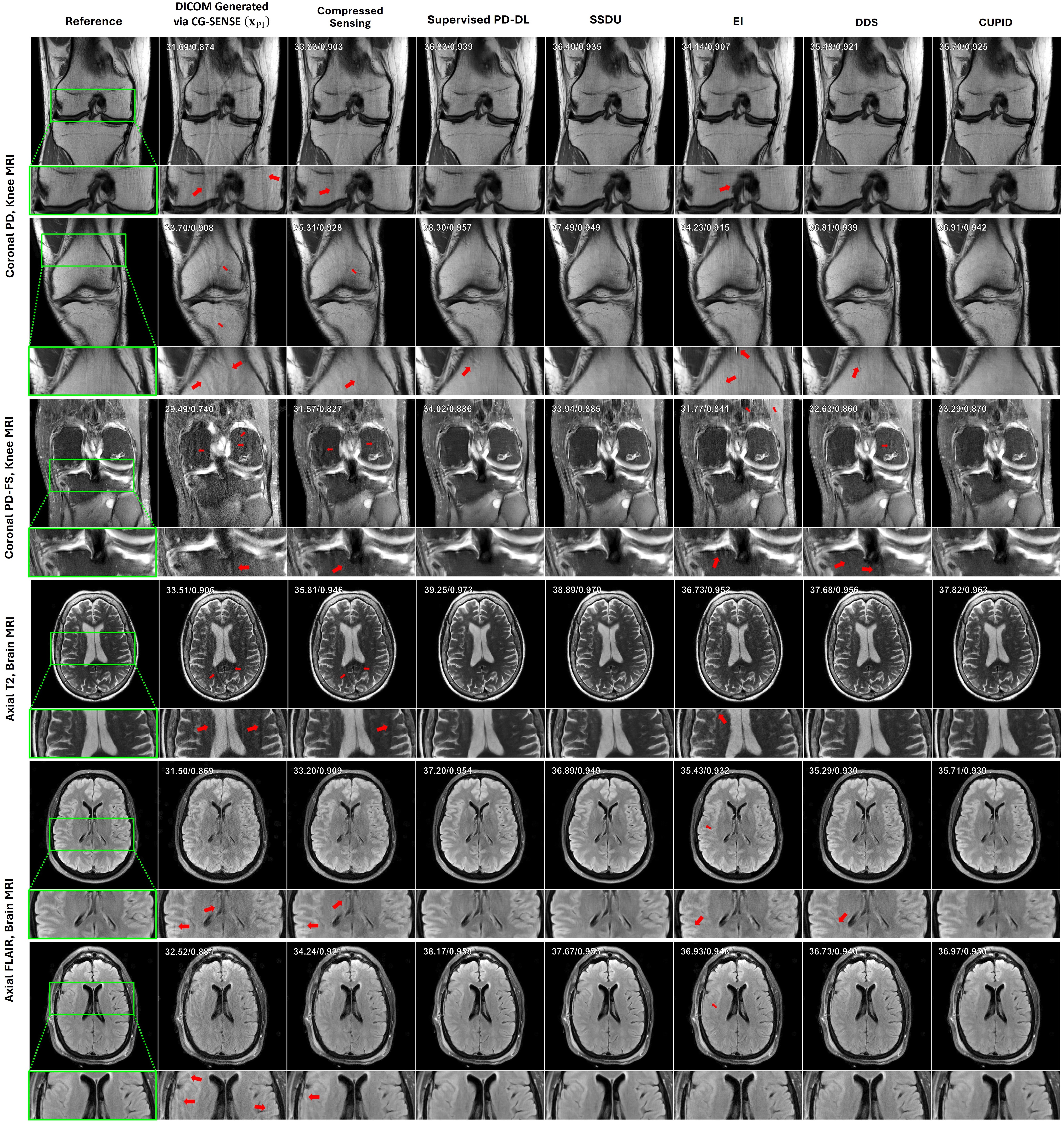}
    \caption{Illustrative slices reconstructed at $R=4$ using equidistant undersampling from coronal PD and coronal PD-FS knee MRI, as well as axial T2 and FLAIR brain MRI. CG-SENSE, CS, EI and DDS exhibit artifacts. On the other hand, CUPID surpasses them with high-fidelity reconstructions, closely matching supervised and self-supervised PD-DL methods that require raw k-space data during training.}
   \label{fig:more-retro}
\end{figure}

We provide additional qualitative reconstruction results for various methods, showcasing representative reconstructions from each dataset, as illustrated in \figref{fig:more-retro}.

\end{document}